\documentclass[%
 reprint,
 amsmath,amssymb,
 aps,
prb,
]{revtex4-2}

\usepackage{graphicx}% Include figure files
\usepackage{dcolumn}% Align table columns on decimal point
\usepackage{bm}% bold math
\usepackage[version=4]{mhchem}
\begin{document}

\preprint{APS/123-QED}

\title{Versatile Domain Mapping Of Scanning Electron Nanobeam Diffraction Datasets Utilising Variational AutoEncoders and Decoder-Assisted Latent-Space Clustering}% Force line breaks with \\

\author{A. Bridger}
\affiliation{%
Inorganic Chemistry Laboratory, University of Oxford, Oxford OX1 3QR, UK
}%
\affiliation{%
ISIS Neutron and Muon Spallation Source, STFC Rutherford Appleton Laboratory, OX11 0QX, UK
}%
\affiliation{%
Electron Physical Science Imaging Centre, Diamond Light Source Ltd., Didcot,UK
}%
\author{W. I. F. David}%
\affiliation{%
ISIS Neutron and Muon Spallation Source, STFC Rutherford Appleton Laboratory, OX11 0QX, UK
}%
\affiliation{%
Inorganic Chemistry Laboratory, University of Oxford, Oxford OX1 3QR, UK
}%

\author{T. J. Wood}
\affiliation{%
ISIS Neutron and Muon Spallation Source, STFC Rutherford Appleton Laboratory, OX11 0QX, UK
}%

\author{M. Danaie}
\affiliation{%
Electron Physical Science Imaging Centre, Diamond Light Source Ltd., Didcot,UK
}%

\author{K. T. Butler}
\affiliation{%
SciML, Scientific Computing Department, STFC Rutherford Appleton Laboratory, Harwell Campus, Didcot,UK
}%

\date{\today}% It is always \today, today,

\begin{abstract}
Advancements in fast electron detectors have enabled the statistically significant sampling of crystal structures on the nanometre scale by means of Scanning Electron Nanobeam Diffraction (SEND). Characterisation of structural similarity across this length scale is key to bridging the gap between what can be understood about local atomic structure (using atomic resolution techniques such as High Resolution Scanning Transmission Electron Microscopy (HR-STEM)) and what can be understood on the macro-scale (using bulk techniques such as powder X-ray and neutron diffraction). The use of SEND technique allows for structural investigation of a broad range of samples, due to the techniques ability to operate with low electron dosage and its tolerance for sample thickness, relative to HR-STEM. This, coupled with the capacity for data collection over a wide areas and the automation of this collection, allows for statistically representative sampling of the microstructure. Also due to these factors, SEND generates large datasets and as a result automated/ semi-automated data processing workflows are required to aid in maximal extraction of useful information. As such, this paper outlines a versatile, data-driven approach for producing domain maps, as well as a statistical approach for assessing their applicability. The production of such domain maps for a dataset can help highlight nuance in the microstructure, as well as improve the manageability of that dataset for further investigation. The workflow outlined utilises a Variational AutoEncoder to identify and learn the sources of variance in the diffraction signal and this, in combination with clustering techniques, is used to produce domain maps for a set of varied example cases. This approach: is agnostic to domain crystallinity; requires no prior knowledge of crystal structure; and does not require the, potentially prohibitive, simulation of a library of appropriate diffraction patterns. 
\end{abstract}

\maketitle

%\tableofcontents

\section{Introduction}
The drive of material science towards improving the physical properties of a given application is universally aided by improving the fidelity with which materials can be characterised, modelled and understood, across length scales, from nanometres to centimetres. Scanning Electron Nanobeam Diffraction (SEND) can offer a fast and convenient approach for collecting structural information on the nanometer scale\cite{Ophus2014}. While a wealth of insight into the bulk structure can be obtained through techniques such as powder X-ray diffraction (PXRD)\cite{Bunaciu2015} and powder neutron diffraction (PND)\cite{Cheetam1977}, and into the immediate local structure using techniques such as High Resolution Scanning Transmission Electron Microscopy (HR-STEM)\cite{James1999}, SEND can provide a solution to bridge these length scales and produce a more complete structural model. SEND, as a technique, is becoming increasingly accessible alongside advances in fast electron detectors\cite{McLaren2020} and the technique lends itself well to microscope automation. This means it can cover areas in the micrometres length scale and hence, while still capturing microstructural detail not accessible via PXRD, provide statistically significant insight into the crystallographic defects and morphology of the phases present in the sample \cite{Ophus2019}.  

Utilising the 4D Scanning Transmission Electron Microscopy (4D-STEM) data collection approach, SEND combines this with a minimised probe convergence semi-angle to obtain distinct Bragg peaks within the diffraction patterns. The advantages that SEND technique offers make it a versatile approach for studying a wide variety of materials. The primary advantages of SEND technique are: relaxation of sample thickness requirements (relative to HR-STEM) resulting in data collection from much wider areas of sample, which in turn offers better statistics of the microstructure; the ability to operate the microscope in a stable configuration, without requiring the frequent aberration tuning common in HR-STEM measurements, which permits for longer and more autonomous data collection. Given the relatively large probe size, resulting from the optical setup in SEND, data collection is normally done at low to medium magnification range. This leads to a reduced electron dose subtended on the sample, making this technique particularly suitable for the study of beam sensitive systems, e.g. organic perovskites and Metal Organic Frameworks (MOFs)\cite{Laulainen2019}\cite{Johnstone2020}. Due to these factors, SEND data acquisition is well suited to the task of domain mapping, which is typically achieved through crystallographic orientation mapping\cite{Brunetti2011, Kobler2013, Gallagher2019, Nalin2020, Ortiz2020}. Domain mapping, broadly speaking, is the task of identifying regions which share distinguishing properties and grouping them together. Producing these domain maps for SEND data allows for aggregation of all the diffraction information available within the entire domain into a single, signal boosted diffraction pattern. Working with these signal boosted patterns rather than the individual patterns is advantageous for studying more subtle features like superstructure and can aid pattern indexing, as the signal boosted patterns typically exhibit additional resolvable Bragg peaks, and improved peak definition. If the atomic structure present within the mapped domains can be identified (through direct treatment of diffraction data or combination with complementary techniques), then the statistical nature of SEND technique could provide a link between local structure and the experimental results of the bulk techniques\cite{Khalinin2021}.

%The primary advantages of SEND technique are: the relaxation of the requirements necessary of a sample, allowing for low electron dose studies\cite{Laulainen2019}\cite{Johnstone2020} (and thus the study of beam sensitive materials) and the use of thicker samples (relative to High Resolution Scanning Transmission Electron Microscopy); data collection from much wider regions of the sample, which in turn offers better statistics of the microstructure; and the ability to operate the microscope in a stable configuration, without requiring constant aberration correction, which permits for longer and more autonomous data collection.

\begin{figure*}[t]
\centering
\includegraphics[width=\linewidth]{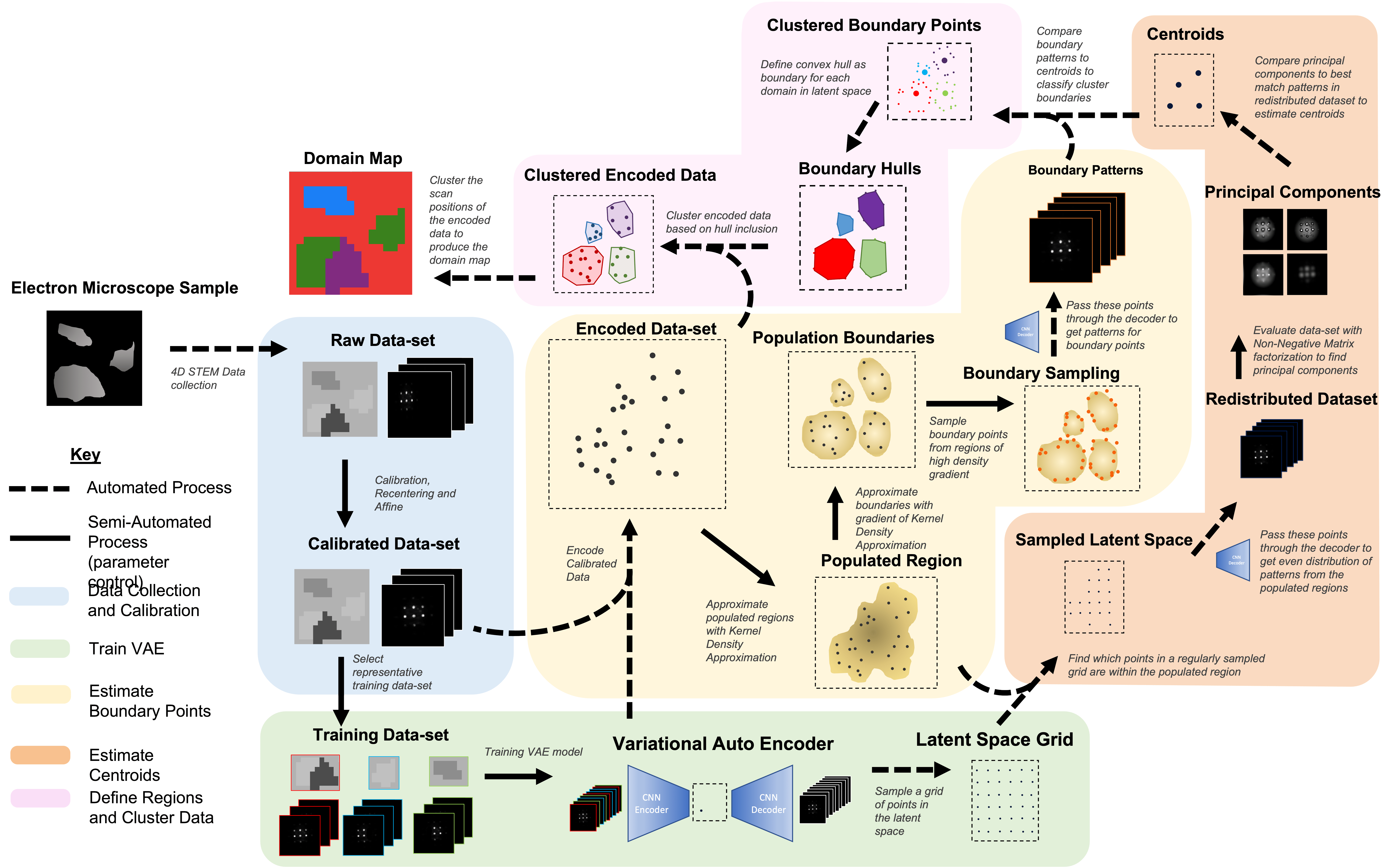}
\caption{Overview of proposed workflow. The motivation of each step can be most straightforwardly understood by working backwards from the final domain map }
\label{fig:workflowflow}
\end{figure*}

For a given dataset, no single, ground-truth domain map exists which is inherent to that sample. Experimentally, there will always be multiple sources of variation between patterns, due to the statistical nature of the sampling. As the domain map seeks to group regions of shared properties, the returned domain map for a sample must depend on what property similarities are relevant to the specific experiment. For a given sample, it is therefore plausible to seek different domain maps, depending on what specific factors are of interest. As an example, given a SEND dataset taken from a thin foil sample with small precipitates dispersed within it, one might only be interested in the locations on the foil where these precipitates have aggregated and the structural and orientation factors present in the matrix that drives precipitation in these locations. Alternatively, for the exact same dataset, one might instead be interested in the precipitate structure, and their orientation relationship with the matrix, in which case, a domain map which identifies differences in bulk/matrix lattice mean crystal structure and relative orientation would be desired. Thus, the goal of the experiment will always enforce bias onto which factors of variance are relevant and this should be considered when separating domains.

Research into automated crystallographic orientation mapping (ACOM)\cite{Ophus2022, Cautaerts2022} has employed SEND data collection to tackle the task of domain mapping. These techniques rely on prior knowledge of possible underlying crystal structures in order to classify the nature of individual diffraction patterns and then aggregate these into maps. 
%%% This seems a little bit too much detail for an intro
For ACOM, a library of possible zone-axis orientations are generated and the individual patterns are correlated to determine the most plausible match. This approach works well when there is prior knowledge of models which can replicate all the desired variations within the experimental diffraction data collected. Unfortunately, this is frequently not the case. In many situations the simulations obtained from the models can differ from the experiment due to factors which are absent, overlooked or incorrect. Example factors such as: nuanced superstructure\cite{Meyer2021, Liu2020}; overlapping lattice motifs\cite{Valery2017}; non-systematic dynamical effects; or simply a lack of structural insight. Non-systematic dynamical effects usually cannot be fitted and so attempts must be made to mitigate them experimentally\cite{Vincent1994}. These simulation approaches are also unable distinguish amorphous regions within the sample from vacuum or thick sample. 

This work reports a structurally agnostic, data-driven workflow for domain mapping, which requires no prior knowledge of possible crystal structures and is equally as effective for identifying amorphous regions with little to no crystalline diffraction character.
Techniques based on feature extraction with non-negative matrix factorisation (NMF)\cite{Bruefach2022} and hierarchical k-means clustering to group similar diffraction signal\cite{Shi2022} have recently been reported.
%Similar, contemporary techniques [cite Alex's paper] utilise curated feature extraction techniques (Bragg disk detection, Angular Averaging and Radial variance) followed by an iterative application of non-negative matrix factorisation (NMF) to produce clusters\cite{Allen2021}. These clusters are merged based on how much they correlation, until they pass below a desired threshold, this is then outputted as the final domain map. While this technique provides a promising approach for producing a domain map without need for prior structural knowledge, we report a complementary technique, driven by similarity metrics between diffraction signals. Other techniques use hierarchical k-means clustering to group similar diffraction signal\cite{Shi2022} for strained material systems, with promising results. 
The workflow, discussed within, leverages a Variational AutoEncoder (VAE) to identify the variance and similarity within the diffraction data. VAEs have recently shown great promise for processing and understanding large scientific datasets, extracting useful representations of data \cite{oviedo2021interpretable, liu2022experimental, allotey2021entropy}. For this workflow, the VAE is used in tandem with NMF to produce domain maps. The  VAE simultaneously improves the signal-to-noise ratio and uses similarities between 2D diffraction patterns to optimally cluster regions in the low dimensional manifold of the VAE latent-space. This approach means that the factors determining the segregation are tangible and interpretable, allowing for simple alignment of the bias in the output domain map with the experimental goals. Additionally, this approach produces a confidence map of the domain assignments, which means that the experimentalist can manually investigate regions of ambiguity and, if desired, update the mapping process.

%By utilising relative positions in the latent-space and decoded, low signal-to-noise diffraction signals, the workflow maintains its primary driver for classification being the similarity between the 2D diffraction signals. 
%This approach is beneficial when considered in the context of domain maps requiring the bias of an experimental to inform them, as the factors driving the domain segregation remain tangible and interpretable, allowing for more simple alignment of the bias in the output domain map with the experimental goals.  

This approach combines the convenience and rigour of automated mapping, with the flexibility of human-in-the-loop steering of the desired outcomes. This report begins by outlining the workflow as shown in Figure \ref{fig:workflowflow}. The paper then demonstrates the procedure, first on a synthetic dataset, which allows for quantifiable comparison with ground truth, then on a diverse set of real experimental datasets. It is shown that this method can produce meaningful domain maps in different scenarios where different applications are required. The code required to run this method is available in an open repository\cite{coderepo} with the intention that the flexibility of this approach will make it a valuable tool for the application of SEND to gain statistically relevant insights into composition-structure-property relationships in materials.

%An overview of the reported workflow is shown in Figure \ref{fig:workflowflow}. As additional benefits, such domain maps can allow for compressed representations of a dataset (which reduces storage requirements), and also can be used to produce improved signal-to-noise ratio patterns by averaging the domains. 

\section{Workflow}
\subsection{Overview}
The input SEND dataset is in the form of a 4-D data array (x,y | u,v): two dimensions corresponding to the x,y coordinates of the probe position in the real space plane; and two dimensions corresponding to u,v coordinate of the pixelated detector recording diffraction data intensities in reciprocal space. This dataset is then pre-processed and used to train a VAE. The VAE is constructed in two parts, encoder and decoder, which are trained in tandem to produce the identity transformation (i.e. take the diffraction signal as input and return the same signal as output), passing through an n-dimensional latent-space (where n $<<$ number of detector pixels). Thus the VAE learns to produce a compressed representation of the data. Once trained, the two halves can be used independently, to convert diffraction signal into low dimensional coordinates (encoder) or low dimensional coordinates into diffraction signal (decoder). 

As standard, this workflow uses two dimensions for its low dimensional coordinates. The properties of both the encoder and decoder are leveraged by this workflow, but in the first case the encoder is used to convert the pre-processed patterns of the input data into a collection of points within a 2-D plane (green section to yellow section in Figure \ref{fig:workflowflow}). As a result of the constraints of the tandem training of the components of the VAE, the 2D representation outputted from the encoder will group patterns of similar diffraction character proximally within this latent-space. Directly clustering these points based on 2D coordinate alone is non-trivial. While similar patterns will be encoded near to each other, a point being nearby is not guaranteed to be from a similar pattern. This can leads to ambiguity in identifying the boundaries between regions from point locations alone. 

In order to most effectively and efficiently cluster patterns into separate domains the workflow uses a four step process: (i) estimate where the boundaries of domain regions could be in the latent-space; (ii) estimate where a representative pattern of each domain is located; (iii) use the Structural Similarity Index (SSI)\cite{Wang2009}\cite{Wang2004} of representative patterns and decoded boundary patterns to define the boundary of each domain's region in the latent-space; and (iv) use the location of encoded data relative to these boundaries to cluster our dataset. Using this workflow, our sample data can quickly be assigned to clusters and these labels can be used to produce domain maps of the sample. We now explore each of these steps in more detail.

\subsection{Variational AutoEncoder}
\label{subsec:vae}
The AutoEncoder was first developed in the late 1980s\cite{Gallinari1987, Lecun1987, Rumelhart1986} and was later extended to the VAE by replacing each single latent-space coordinate with a pair of values which define the statistical distribution from which that coordinate could be sampled \cite{Dayan1995}. 
%This is done as a means to add regularisation to the latent-space and create a smoother distribution of embedded points. 
The function of the VAE within our workflow is to act as a flexible approach for identifying the most important sources of variation within the diffraction dataset and embedding these patterns conveniently into a low dimensional representation.

\begin{figure}[htbp]
\centering
\includegraphics[width=\linewidth]{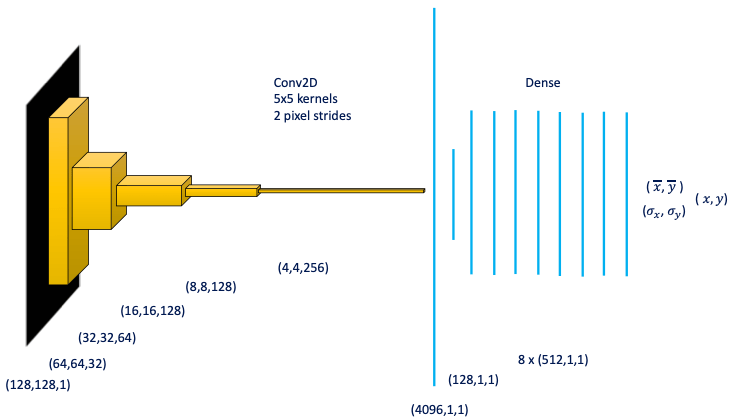}
\caption{Architecture of Encoder half of VAE, the decoder is the same model structure in reverse}
\label{fig:modelarch}
\end{figure}

The VAE was built using Tensorflow \cite{Tensorflow} and the model architecture is summarised in Figure \ref{fig:modelarch}. The trained models are also available in a repository associated with this paper \cite{coderepo}.

Upon training, the VAE has no inclination as to the physical phenomena underlying the observed data nor an appreciation of what aspects of the dataset one might be interested in. As the VAE trains it seeks to minimise the loss function which is calculated as the sum of mean squared error (difference between input and reconstructed patterns) and Kullback-Leibler Divergence\cite{Kullback1951} (deviation of the distributions in the latent-space from the normal distribution). The loss is minimised by modifying the parameters which control the encoding into the latent-space and the decoding of these values to a reconstructed pattern. 
%As the latent coordinates are statistically sampled the model favours embedding patterns which will return similar results, under the loss function, nearby. 
Performance is biased towards optimising the separation of the largest sources of variance within the training dataset first 
%(as these improvements yield the largest gradient descent in parameter optimisation) 
 and then refining the model for smaller and smaller sources of variance. Both these biases define how the model will separate out the input dataset and essentially defines the model's understanding of similarity. 
 
In order to use the VAE most effectively, it is therefore necessary to manipulate these biases to align the VAE's interpretation of similarity with the definition of similarity imposed by the focus of the particular experiment. This is most effectively achieved by modifying the training set to contain the representative proportions of the relevant sources of variance and truncating training once all the desired variance has been identified. This truncation is achieved by saving iterations along the training process and evaluating the performance at different stages. This evaluation can be done by visually comparing the patterns inputted and outputted by the model and also by comparing decoded points from the different regions in the latent-space to input the patterns that have been encoded into those regions. If the bias is controlled effectively, the VAE can be used in vastly disparate use cases to give a primary identification of which patterns fall within which domain.

\subsection{Latent Clustering}
In order to cluster the patterns into single domains, the 2D encodings are used in tandem with the decoder and the Structural Similarity Index (SSI)\cite{Wang2009}\cite{Wang2004}. The aim is to find and define the regions within the 2D latent-space, that correspond to each domain. To achieve this, the workflow seeks to identify the locations of the all boundaries between these domain regions, and also identify a typical pattern for each domain (yellow and orange sections of Figure \ref{fig:workflowflow}, respectively). Then, by passing points along the boundaries through the decoder to generate diffraction patterns, the workflow uses the SSI image similarity metric between these decoded patterns and the domain typical patterns, to identify which boundaries points correspond to each domain.

\subsubsection{Boundary Point Estimation}
\begin{figure}[htbp]
\centering
\includegraphics[width=\linewidth]{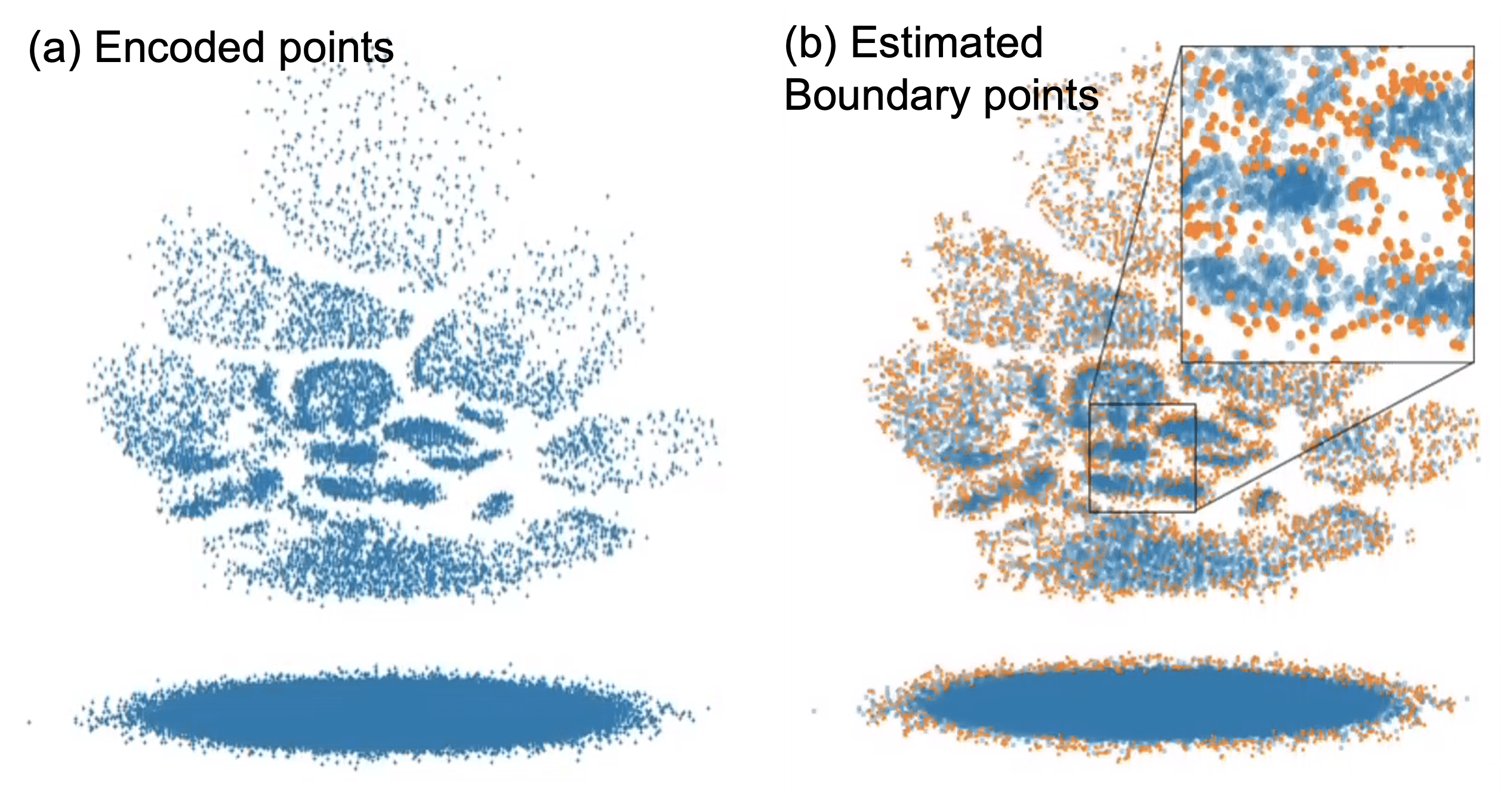}
\caption{(a) 96,000 diffraction patterns encoded into 2D space. (b) the same encoded patterns in blue, with 5,000 sampled boundary point candidates shown in orange. A sub-region is highlighted in the inset}
\label{fig:boundaryest}
\end{figure}

The first requirement for the clustering approach is to identify candidate points that could define the boundaries of each domain. It is assumed, due to the bias of the VAE, that the encoded data will be grouped such that similar patterns are nearby and so each domain will occupy a continuous, definable region, within the latent-space. 
%The robustness of this separation depends on how aligned the bias of the VAE is with the experiment, as well as the inherent factors of each domain, such as population size and variance. That is to say, 
The density of 2D points within domains and between domains can vary across the latent-space. As such, density-based evaluation is sufficient to identify probable locations of domain boundaries (typically areas where there is large change in point density), which makes it fit for the desired role of approximating the boundaries (although it rarely identifies these areas exclusively, this is not a problem). Specifically, the density-based evaluation is achieved by using a Kernel Density Estimation (KDE). The KDE is performed using built-in functions in sklearn\cite{Brucher2011} and a brief overview is given in Section \ref{sec:kde}. The KDE is used to approximate the density function for the entire latent-space. The first derivative of this function is then computed and values within a set range are used to define the probability distribution function from which the potential boundary points can be sampled. An example result of this is shown in Figure \ref{fig:boundaryest} 

\subsubsection{Centroid Estimation}
The aim is to use the sampled candidate boundary points to define a bounded area for each domain region. To achieve this, the approach is to identify a set of patterns (referred to as centroids), one for each domain, such that the decoded pattern of a general point within a given domain's boundary, will give a larger SSI when compared to that domain's centroid than any of the other centroids. Due to the continuous nature of the latent-space, almost any point within the region should fulfil this role. Thus the centroids can be estimated by a few approaches: using the derivatives of KDE to find stationary points in the density function which correspond to centres of boundary regions; being manually selected; or by using NMF \cite{Hien2015}. NMF is the most automated approach and is used as standard in this workflow. 

%or by using Principal Component Analysis (PCA)\cite{Pearson1901}. PCA is the most automated approach and is used as standard in this workflow. 

In order to find a set of patterns to be representative of the domains within the latent-space by NMF, the workflow aims to use a dataset of diffraction patterns that contains all the pattern variation which has been encoded within the latent-space. Then the NMF is applied to extract the different sources of variations in the patterns that are found within this dataset. Assuming that the patterns within each domain exhibit some characteristic of their pattern that distinguishes them from the other domains, then these characteristics are sources of variation within the dataset. As such, these characteristics will be identified by the NMF. As each characteristic will be shared by the patterns within this domain but not in the other domains it should fufill the role of the centroid, as defined earlier. 

The dataset used could just be the original data, but this will have performance issues due to the uneven sampling of domains. The original data will be larger than is required, due to oversampling points in the more populated domains, and so will take longer to perform the NMF. Further, as the less populated domains will be significantly undersampled, the source of pattern variation corresponding to these might prove harder to extract from the inherent noise within the more populated domains. 

To generate a better performing dataset for centroid estimation, the workflow, evenly samples points across the populated regions of the latent-space, using the already calculated density approximation to identify the populated regions. These points can be fed through the decoder to generate a collection of diffraction images representative of the pattern variance within the original dataset. This re-balanced dataset reduces the issues identified for the raw dataset. Firstly, it is a much smaller dataset while still containing all the variance the VAE has identified within training; this allows for speeds up in calculation of the NMF. It is also a significantly de-noised dataset, as it only contains the variances that have been identified as relevant during the VAE training. Removing the sources of variation not corresponding to changes in domain, will improve the applicability of these NMF components for use as centroids. Finally, the distribution of patterns within the dataset is less dependent on the occurrence frequency within the sample and will be proportional to the size of region in the latent-space. This increases the visibility of minor phases and in turn results in better the segmentations of less populated domains.

An NMF is performed on the re-balanced dataset. The workflow then compare these components to the patterns from the rebalanced dataset and replace the NMF component with the reconstructed pattern that most closely matches it, as measured by SSI. This is because the NMF component will contain only the distinguishing characteristic of that pattern, but ideally the centroid would be the entire diffraction pattern that is representative of that domain, not just part of it. This set of patterns will then correspond to the centroids of our domains. 
The total number of these centroids is therefore dependent on the number of components requested of the NMF. The fewer the components requested of the NMF, the larger the sources of variance in the dataset has to be in order to be classified as a component. To reduce the dependence of the clustering on this parameter, the workflow allows for comparisons and evaluation of domain maps using different level number of centroids. Additionally, once the centroid patterns have been identified by NMF, they can be compared by SSI and patterns centroids which are sufficiently similar (by a user-defined threshold) can be merged (by averaging the point locations in the latent-space). By doing this, the number of domains is slightly more flexible and is driven by a maximum level of similarity between domains, rather than being directly determined by a chosen parameter.

\subsubsection{Assigning data utilising convex hulls}
Once the centroids and potential boundary points have been identified, the SSI can be used to assign each boundary point to the best matching centroid. Then, for each centroid, the classified boundary points can be used to define an outer limit of that domain region using a convex hull. The convex hull is used as it allows for faster evaluation of points falling within the hull. A convex hull is applicable as despite not producing a perfect bounding hull, the inaccuracy should only produce false positive inclusions, which should then lead to multiple group inclusion conflicts which can be straightforwardly resolved. With the latent-space now defined by these hulls, encoded patterns can fall into one of three categories: not within any hull; within exactly one hull; within more than one hull. The vast majority of the points tend to fall within exactly one hull. The points shared between hulls are classified by direct SSI of the pattern with the conflicting hull centroids and the points not within any hulls are classified by direct comparison with all hull centroids.

\subsubsection{Iterating}
Once the points have all been classified into domains, an additional means of centroid approximation becomes trivial, taking the centre of mass (CoM) of the encoded points within each domain. This can offer a useful approach for refining a centroid location and might improve the outcome of a successive iteration of the workflow. This can be beneficial if the region is approximately centrosymmetric and the CoM approximation moves the centroid more centrally within the region. As standard the workflow outputs the CoM of the identified region as well as the centroid used. 

%This approach of defining regions by Convex Hulls speeds up the assignment compared to direct comparison significantly, as the boundary points can be far more sparsely sampled than the total number of patterns. This was seen with one dataset of size 511x511 (261,121 patterns) which was clustered using 5,000 boundary points in 3 minutes, compared to the direct comparison of the patterns to the centroids which took 59 minutes on the same hardware.

\section{Case Studies}
In this section, three case studies are presented to show the diverse use-cases for the workflow. These case studies are: a dataset populated with simulated diffraction patterns for known, distinct orientations of a single crystal structure; an experimental SEND dataset from an Al-alloy thin foil with precipitates; and an experimental dataset of a particle containing a mixture of phases and orientations of layered metal oxides (LMO). 

\subsection{Simulated dataset}
\begin{figure}[htbp]
\centering
\includegraphics[width=\linewidth]{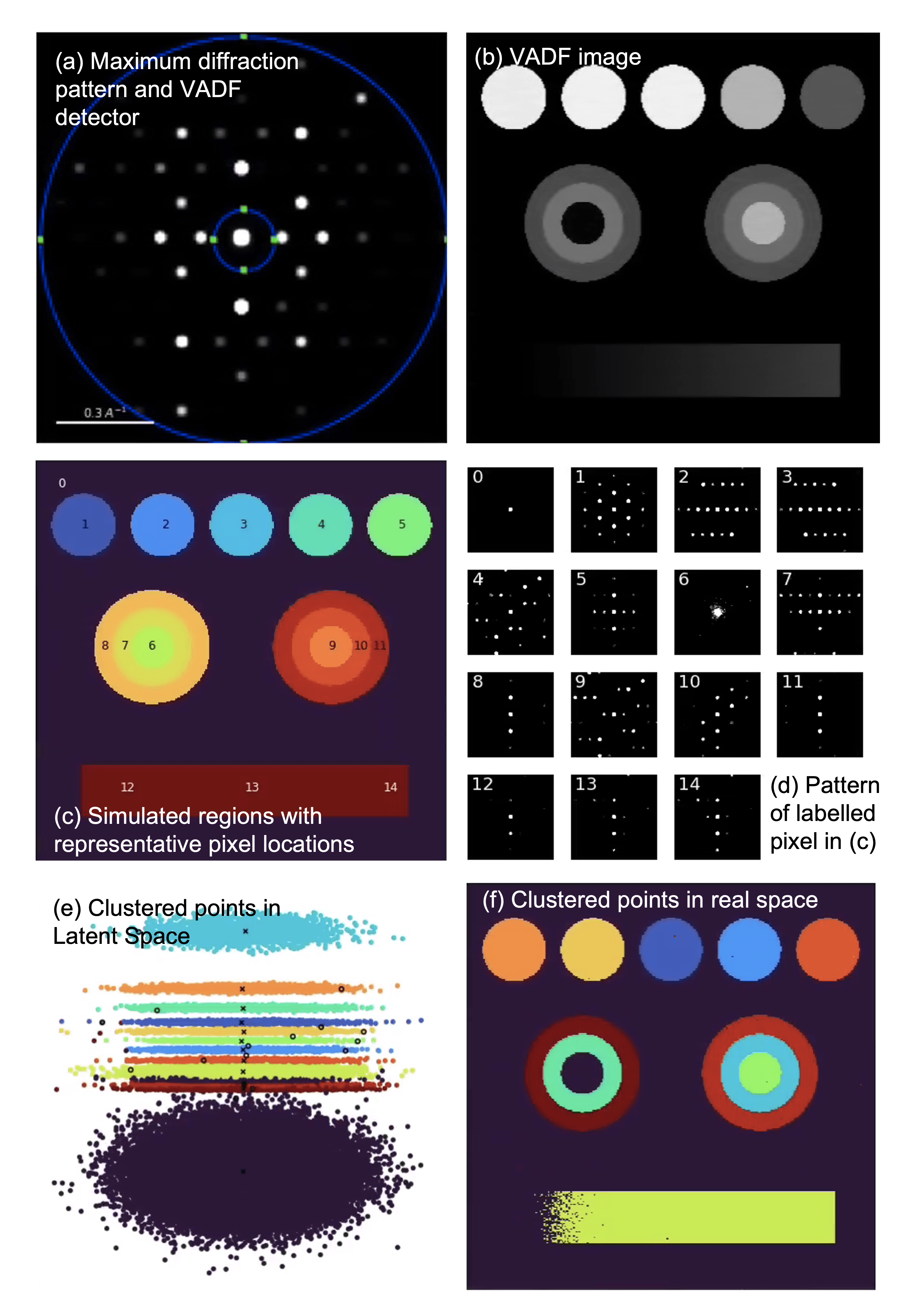}
\caption{(a) the maximum diffraction pattern of the entire dataset with a virtual annular detector. (b) the Virtual Annular Dark Field (VADF) image for the simulated Layered Metal Oxide (LMO) sample dataset. (c) The region map used to populate the simulated dataset, each region contains patterns of the same crystal structure (or amorphous structure) and orientation but with added noise (and variation in thickness for the bar at the bottom). Pixel locations have been indicated with numbers to aid the understanding of the patterns which populate the dataset (d) the diffraction pattern corresponding to the pixel locations highlighted in (c). (e) the encoded positions of the patterns clustered by colour. The open black circles show the initial centroids identified by NMF and refined by SSI. The black crosses show the new centroids by averaging the locations of points within clusters. (f) the clustering on the real space pixel locations, a pixel location has been highlighted in each of the identified domains.}
\label{fig:simdatafull}
\end{figure}

The simulated dataset is populated using 12 distinct zone axis orientations of an LMO structure (as detailed in \ref{subsec:simdatadetails}) and is shown in Figure \ref{fig:simdatafull}(c). The regions 1--5 and 7--11 are each populated with the simulated pattern of a single zone axis orientation. The bar at the bottom containing labels 12, 13 and 14 is populated with the simulated pattern of a single zone axis, but the diffraction signal decreases in intensity from right to left, this is designed to emulate real world conditions where signal intensity might be reduced due to factors such as thickness changes or beam damage. Region 7 contains patterns for an amorphous region. The final region, 0, is populated with an unscattered electron probe to emulate the vacuum around particles of interest. The goal of the domain mapping for this dataset was to identify the areas of different zone axis orientation.

The dataset was generated with a $296 \times 295$ real space and $128 \times 128$ pixel array in each diffraction pattern. This pattern size is ideal for training the VAE and no pre-processing was required. The model was trained, from scratch, on all the shuffled input patterns with a random sample of 1000 patterns used to monitor performance. Training was truncated at the point where the model began identifying distinctions that were deemed no longer relevant to the aim of the segmentation. Once the best iteration of the model is selected the dataset is processed by the workflow and the results are shown in Figure \ref{fig:simdatafull}(e) and (f).

Comparing the output to the ground truth the consistency of the clustering is above 98\% for all regions excluding the intensity varying bar, which identified 86\% of the total patterns. The drop in performance for the bar is a result of patterns with very low diffraction signal that closely resemble the vacuum. This resemblance grows increasingly along the bar such that at the far left the diffraction signal is far more akin to vacuum than the actual structure present, so this could be considered a valid interpretation of the domains. 

\subsection{Precipitates in Al alloy thin foil}
\label{HMA-sec}

Precipitation-hardening is an important mechanism for increasing the strength of commercial aluminium alloys. Gaining statistical insight into the population and crystallographic nature of the precipitates in these alloys is experimentally challenging and thus SEND is a suitable technique to tackle this problem\cite{Sunde2020}. The experimental dataset presented here is an example of such SEND data. This dataset is comprised of $511 \times 511$ probe positions in real space with $256 \times 256$ pixel array in diffraction patterns. The goal of the segmentation is to identify the diffraction character of the precipitates, which form a minor component of a dataset dominated by the diffraction of the Al substrate.

To optimise the training for this functionality the data is pre-processsed to crop to the central $128 \times 128$ of the diffraction pattern, where the majority of the desired information can be observed. The training data is then re-sampled using the Shannon entropy\cite{Shannon1948} of the processed patterns as an indicator of precipitate character. The Shannon entropy is a measure of information content in an image and is given by $-\sum_k p_kln(p_k)$, where $p_k$ is frequency/probability of pixels of value k. Shannon entropy is used as the precipitate patterns are expected to have a higher entropy due to the additional diffraction information and is based on a similar application of Shannon entropy by Slouf et. al.\cite{Slouf2021}
%The VAE was trained and binary clustering was used to create a mask of precipitate regions.
The VAE was trained and the resulting latent-space clustered into two regions, which correspond to the precipitates and the background Al matrix. 

\begin{figure}[htbp]
\centering
\includegraphics[width=\linewidth]{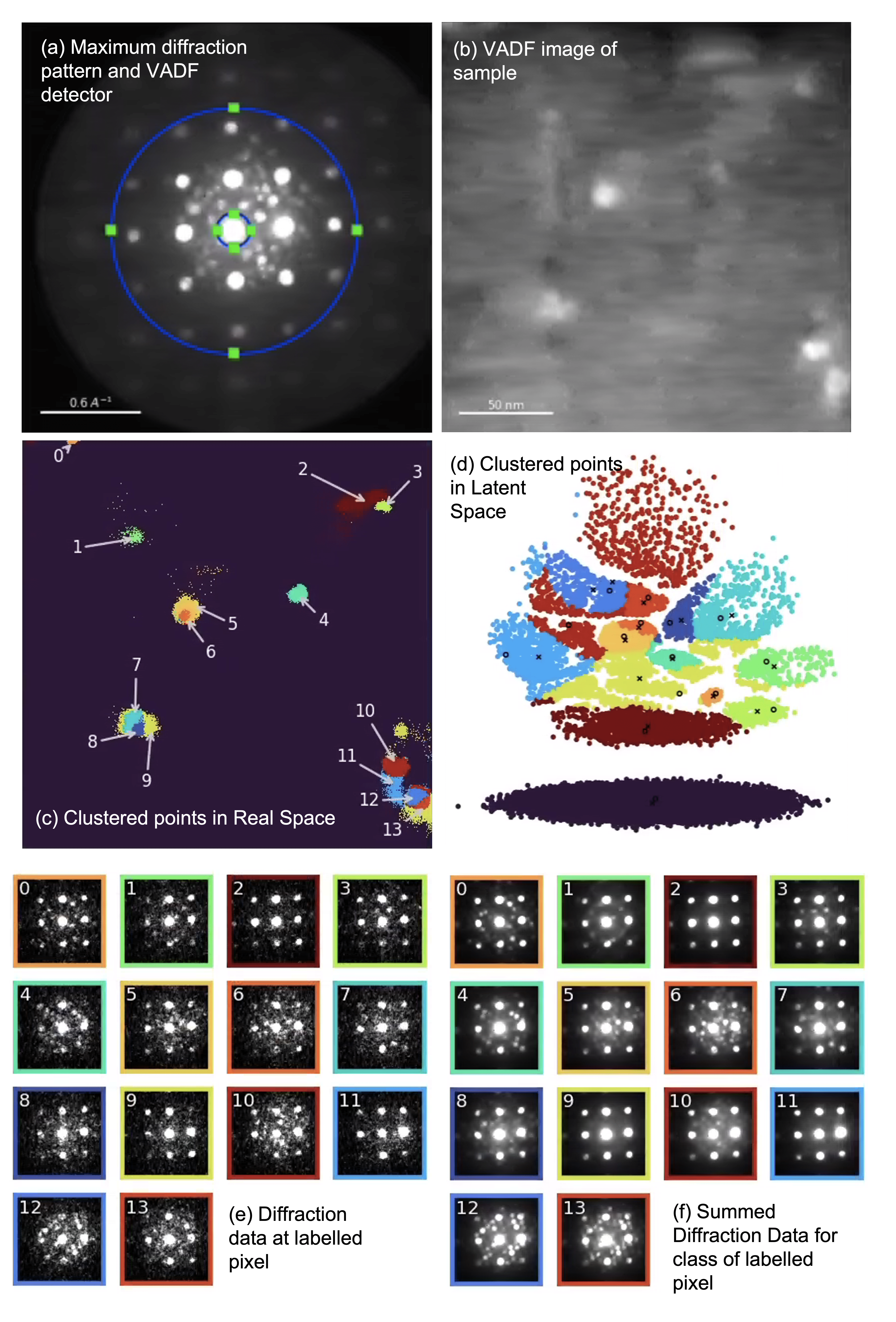}
\caption{(a) the maximum diffraction pattern of the entire dataset with a virtual annular detector. (b) the Virtual Annular Dark Field (VADF) image for the precipitates in Al alloy thin foil dataset. (c) the clustering on the real space pixel locations, a pixel location has been highlighted in each of the identified domains. (d) the encoded positions of the patterns clustered by colour. The open black circles show the initial centroids identified by NMF and refined by SSI. The black crosses show the new centroids by averaging the locations of points within clusters. (e) The single diffraction pattern corresponding to the pixel highlighted in (c), the pattern border colour corresponds to the colour of the class in the segmentation. (f) The mean diffraction pattern corresponding to the class of the pixel highlighted in (c), the pattern border colour corresponds to the colour of the class in the segmentation.}
\label{fig:hmafit}
\end{figure}

Other methods for masking the precipitate region will yield similar results more straighforwardly, but this workflow was used to illustrate how VAE bias can be directed using iteration and how this might be applicable in less trivial cases where classical masking techniques aren't applicable. The mask was used to generate a new 4D dataset with the variance of the background Al substrate replaced with a single mean pattern. This new dataset was then used to retrain the VAE and the model used to cluster the navigation region using 16 components in the NMF as indicated in Figure \ref{fig:hmafit}.

From the clustering in Figure \ref{fig:hmafit}(d), it can be observed that the boundaries of the classes do not precisely follow the gaps between the more dense point encodings (the yellow-green class can be seen to span across various point densities). This is a similar case to that of the thickness effects for the simulated data, as again there is deviation from where one might anticipate boundaries, but again, this is the result of regions with insignificantly low diffraction intensity merging. Inspection of the resulting patterns in Figure \ref{fig:hmafit}(e) and (f), corroborates this observation. Separating out the low diffraction information regions from the strongly reflecting centres is an entirely valid domain distinction in this use case.

From Figure \ref{fig:hmafit}, the utility of this workflow can be seen. The workflow has taken a sample of 261,121 diffraction patterns and condensed all the information relevant to the experiment into a domain map and 14 patterns. Additionally, in almost all the examples the signal-boosted pattern for each domain contains better resolved and more usable diffraction information than in the individual patterns.

\subsection{Phase Distribution in Layered Metal Oxides Powder Sample}
\label{p2-sec}

\begin{figure}[htbp]
\centering
\includegraphics[width=\linewidth]{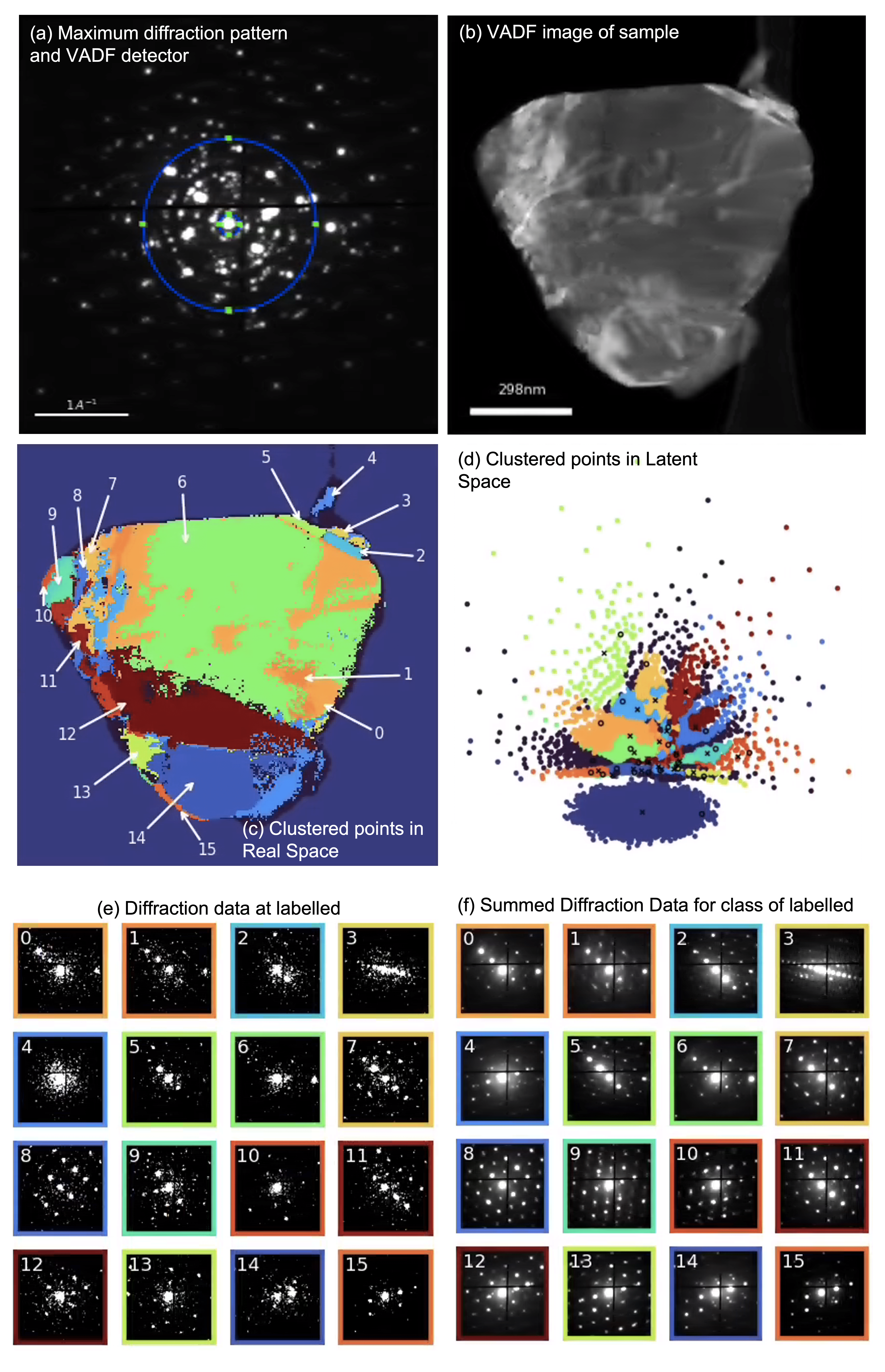}
\caption{(a) the maximum diffraction pattern of the entire dataset with a virtual annular detector.(b) the Virtual Annular Dark Field (VADF) image for the Layered Metal Oxide dataset. (c) the clustering on the real space pixel locations, a pixel location has been highlighted in each of the identified domains. (d) the encoded positions of the patterns clustered by colour. The open black circles show the initial centroids identified by NMF and refined by SSI. The black crosses show the new centroids by averaging the locations of points within clusters. (e) The single diffraction pattern corresponding to the pixel highlighted in (c), the pattern border colour corresponds to the colour of the class in the segmentation. (f) The mean diffraction pattern corresponding to the class of the pixel highlighted in (c), the pattern border colour corresponds to the colour of the class in the segmentation.}
\label{fig:p2fit}
\end{figure}

The final case study is a layered metal oxide sample used as cathode material in sodium ion batteries. For these materials, the structural composition can vary across the sample due to local changes in stoichiometry. All the structure types are based on alternating layers of edge-sharing transition metal-oxygen (TM-O$_{6}$) octahedra and layers of intercalated sodium ions. The dataset is comprised of $256 \times 256$ probe positions in real space with $512 \times 512$ pixel array in diffraction plane. The goal of the segmentation is to identify different phases present within the sample, as well as variation in orientation or composition within these phases. For pre-processing, the diffraction information is centred and cropped to $256 \times 256$ and then down-sampled by a factor of 2 to get $128 \times 128$ input patterns. As before, the irrelevant vacuum and carbon support are masked out to remove unnecessary variance from the VAE training. This is achieved by the more straightforward approach than in the previous example: by masking the central beam and selecting patterns with low sum-total intensities. The remaining patterns are again sampled using the Shannon Entropy. The new training data is used to train the VAE and the workflow implemented to produce the domain map in Figure \ref{fig:p2fit}.

It can be seen from the latent-space and the subsequent domain map that this is a more structurally rich sample than the previously studied examples but the workflow provides a robust segmentation, identifying minor phases in classes of fewer than 70 pixels alongside major classes with as many as 10,000 pixels. The segmentation breaks the sample down into 33 largely distinct regions. While some of these regions correspond to multiple classes containing similar diffraction character, this is a conscious decision for the dataset to search for minor factors that might arise due to changes in composition such as super-lattice reflections, and if desired we could tune this by altering the SSI threshold.  

Comparing patterns in Figure \ref{fig:hmafit} to Figure \ref{fig:p2fit} the difference in challenge between the Al alloy with precipitates sample and the layered metal oxide sample can be seen. The individual patterns of the LMO dataset show better signal-to-noise than the precipitate dataset. As a result, there is less of a stark contrast in the domain patterns compared with the individuals, with regards to signal boosting. However, the quality of the domain mapping can still be qualified by the domain patterns presenting the same sets of reflections that are present in the individual. By this assessment, the clustering is successful in creating clusters where the deviation between the mean patterns and the individual patterns are small. There is also evidence of the level of variation within a phase being extracted, with patterns of the same phase with small intensity shifts (such as the row of reflections in 4 and 5) being separated out. 

\section{Evaluating Confidence in Domain Identification}
As discussed in previous sections the exact domain map outputted is primarily dependent on the VAE model and the cluster centroids. Altering the training dataset and the length of training will alter the prominence of the different sources of pattern variation in the latent-space and this broadly defines the variation within the diffraction data that the VAE is able to differentiate. How finely/coarsely these variations are separated into different domains will then be determined by the number of cluster centroids and their locations within the latent-space. As such, altering any of the above factors can lead to a change in the final domain map. In order to increase the confidence that the best domain map is being used to evaluate the data (one with the most experimentally applicable separation of domains) a statistical approach was applied. 

The motivation of a statistical approach is that while the ideal evaluation would be to go through each individual pattern, within a domain, and check that it has been classified correctly, this is untenable for most domains due to to sheer number of patterns. Instead, using a statistical approach, the pixels that are most likely to change class can be identified and evaluation can be focussed on the classification of these patterns. Pixels likely to change class are determined by comparing maps generated with different parameterisations, and finding which pixels are consistently grouped together. 

\begin{figure}[htbp]
\centering
\includegraphics[width=\linewidth]{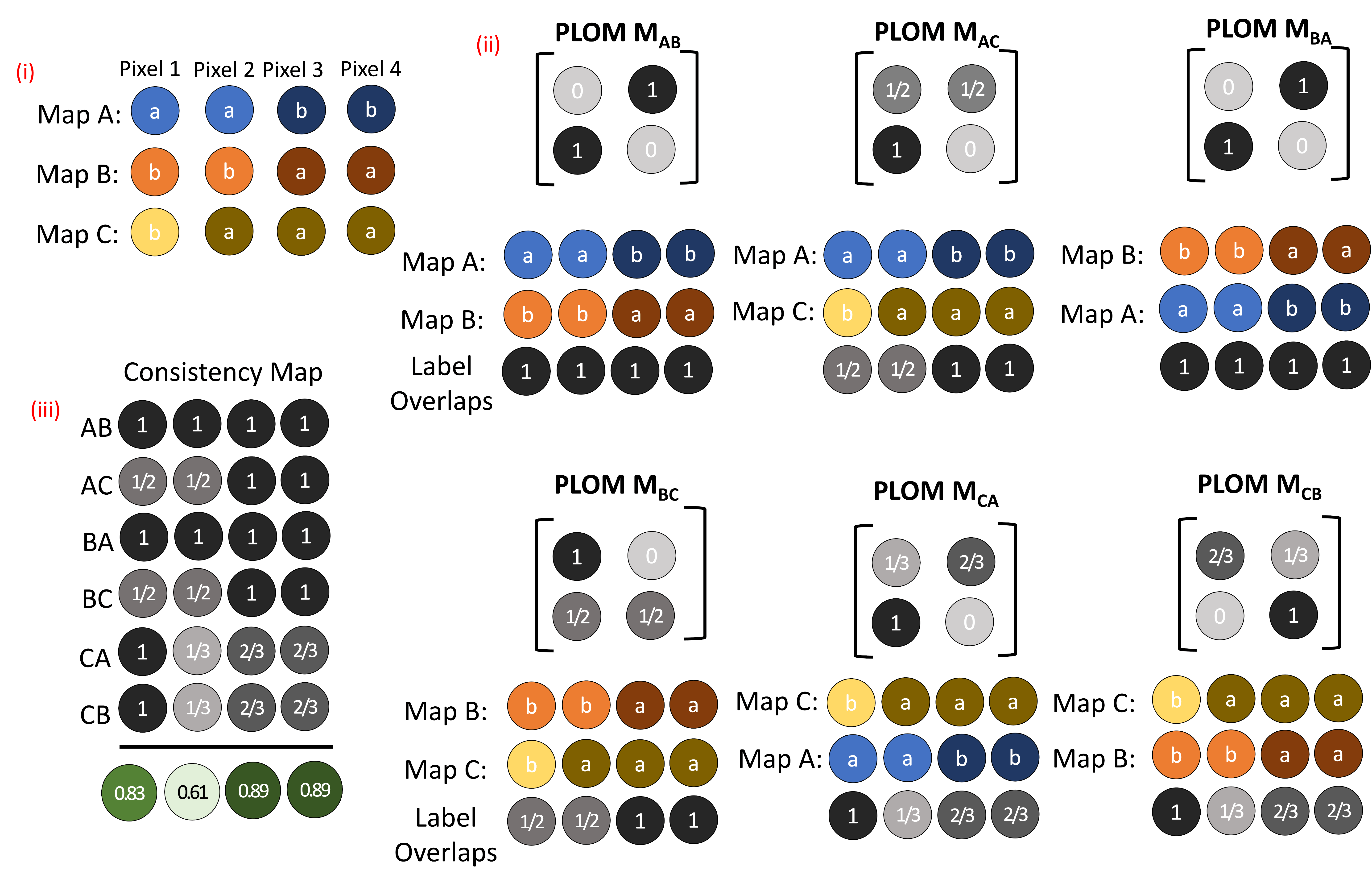}
\caption{An example of the procedure for calculating the Statistical Consistency Map using 3 maps: Map A; Map B; and Map C (denoted by (i)). All the permutations of Ground Truth Map (the top row of each triplet of Map X, Map Y, Label Overlaps) and Comparison Map (the middle row of each triplet of Map X, Map Y, Label Overlaps) are given by (ii) along with the respective PLOM. For each pixel, the proportional label overlap is evaluated between the label of that pixel in the Ground Truth Map and the label of that pixel in the Comparison Map (this can be done by simply referencing the appropriate element of the PLOM).  The Label Overlaps for each of the Ground Truth Map-Comparison Map pairs are then averaged in (iii) to give the final Statistical Consistency Map.}
\label{fig:statmap}
\end{figure}

The consistency is formalised by comparing all pairs of maps (where one is assumed to be the Ground Truth Map and the other a Comparison) and evaluating for each pixel and its label: what proportion of the total pixels sharing this label in the Ground Truth Map also share their labels in the Comparison Map. These proportions for each pair of Ground Truth and Comparison Map are then averaged to give the overall consistency map. For a given pair of Ground Truth and Comparison Map, the proportion of the total pixels sharing one arbitrary label in the Ground Truth Map and another arbitrary label in the Comparison Map can be obtained from the Proportional Label Overlap Matrices (PLOM).
The procedure for calculating a Statistical Consistency Map is summarised below and these steps are illustrated within Figure \ref{fig:statmap}:
\begin{enumerate}
    \item Calculate all Proportional Label Overlap Matrices (PLOM) for every Ground-Truth/ Comparison Map pair
    \item For each pixel in real space:
        \begin{enumerate}
            \item Find the pixel label in the Ground-Truth Map and the label in the Comparison Map
            \item for each Ground-Truth/ Comparison Map pair lookup the proportional overlaps for these two labels from the relevant PLOM
        \end{enumerate}
    \item Average the proportional overlaps of the pixels from one Ground-Truth/ Comparison Map pair over all Ground-Truth/ Comparison Map pairs
\end{enumerate}

\begin{figure}[htbp]
\centering
\includegraphics[width=\linewidth]{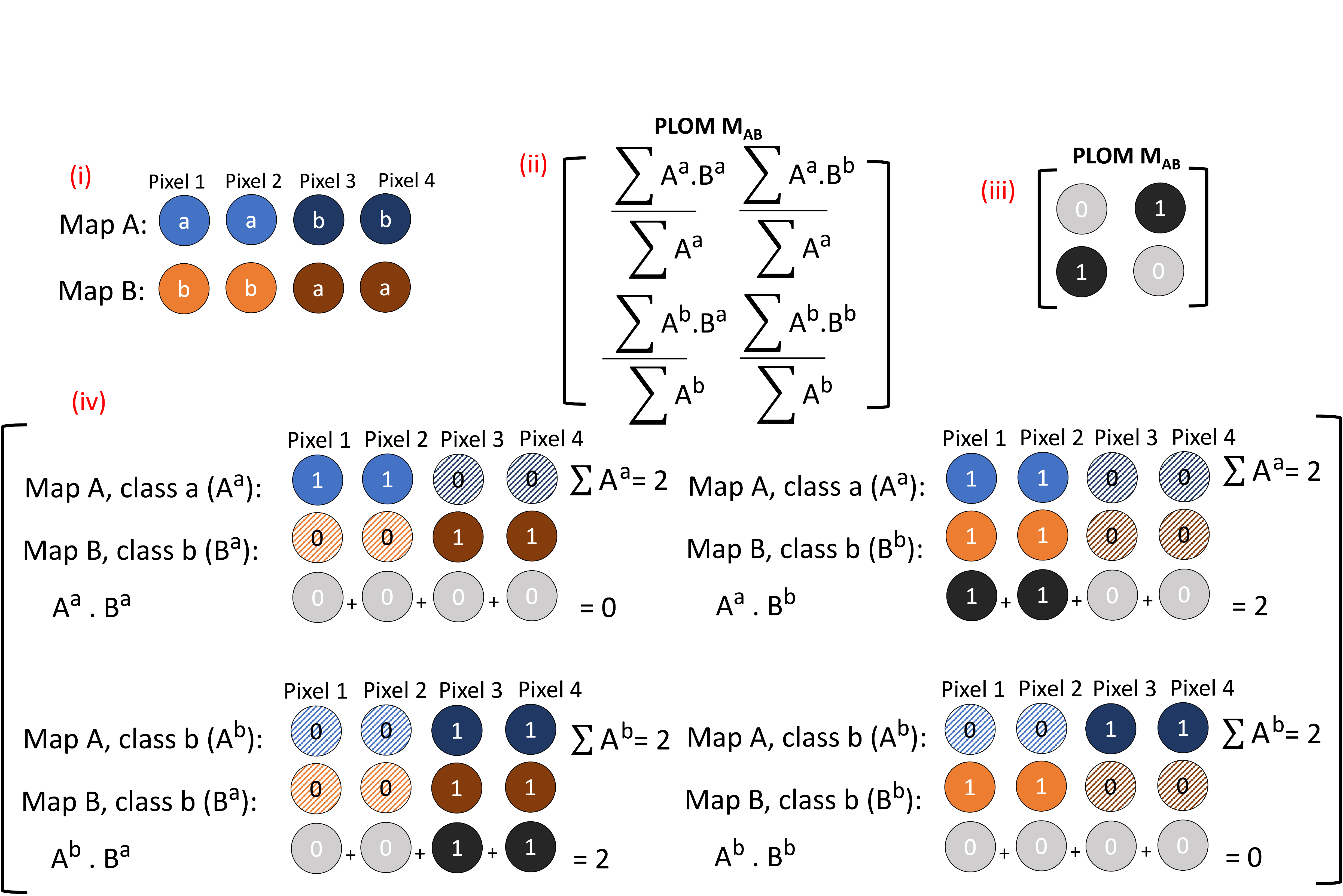}
\caption{An example of the procedure for generating the Proportional Label Overlap Matrix for two 4x1 pixel maps, Map A and Map B (denoted by the red (i), to avoid confusion).  Each map contains two arbitrarily assigned labels, a and b. The matrix marked (ii) has each element populated by the specific calculation required to evaluate that element. The values resultant from these calculations are given in (iii). The intermediate steps for evaluating (iii) are given in (iv) with each element showing: the binary masks for the respective labels of A and B; the total number of the specified A label; and the summation that yields the dot product of the binary masks.}
\label{fig:plom}
\end{figure}

Calculating the Proportional Label Overlap Matrix for Ground-Truth Map A and Comparison Map B is visualised in Figure \ref{fig:plom} and is done using Equations \ref{PLOM} and \ref{binarymask}: where $M^{AB}{_{ij}}$ is the element $ij$ of the PLOM for GTDM Map A and comparison Map B; $A^{i}$ is the Binary mask of label $i$ in Map $A$; and $A_{xy}$ is the class label of the pixel at coordinates $(x,y)$ in Map $A$.
\begin{equation}
\label{PLOM}
M^{AB}{_{ij}} = \frac{A^{i} . B^{j}}{\sum A^{i}}
\end{equation}

\begin{equation}
\label{binarymask}
A^{i} =
    \begin{cases}
            1, &         \text{where } A_{xy} =i,\\
            0, &         \text{where } A_{xy}\neq i.
    \end{cases}
\end{equation}

%This approach takes a set of maps which have been generated with different parameters. Each map in this set in turn is assumed to be the perfect ground truth domain map (GTDM). For each cluster class ($c_{i}$) in the GTDM, each pixel is inspected across each of the other maps in the set. For each pixel, the class label for that pixel ($c_{j}$) in each of the other maps is identified, and the proportion of pixels shared by that class ($c_{j}$) and the class in the GTDM ($c_{i}$) is calculated. This is then repeated for all maps and the proportions are averaged to give an overall consistency value for that pixel. This can then be repeated for all the pixels to produce a consistency map for the dataset.%

\begin{figure}[htbp]
    \centering
    \includegraphics[width=\linewidth]{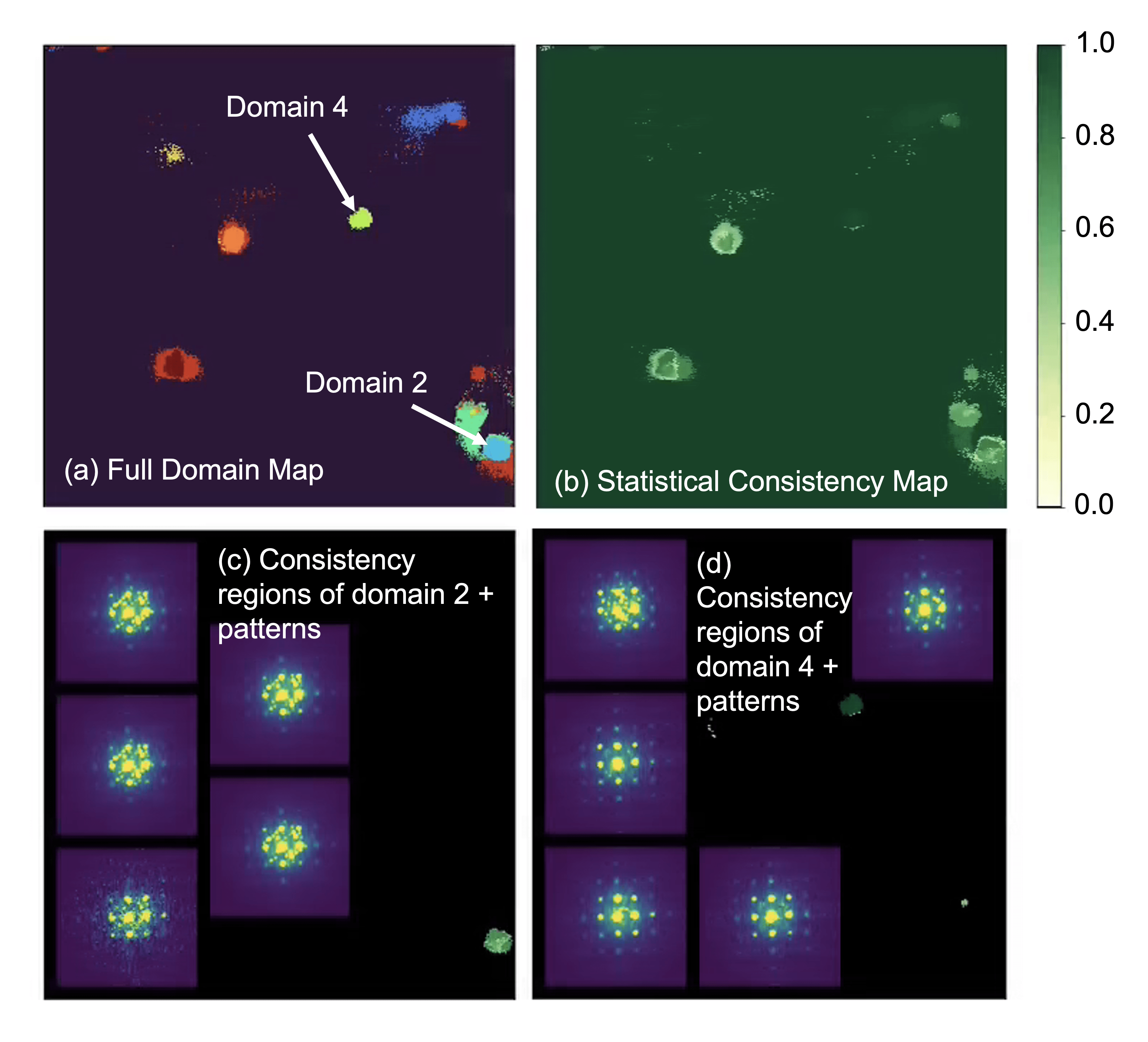}
    \caption{(a) domain mapping of Precipitates on Al alloy dataset using 10 components for centroid estimation. (b) Statistical Consistency Map for the domain mapping of Precipitates on Al alloy dataset, comparing 8 domain maps generated using 10,12,14,16,18,20,22 and 24 principal components for the centroid estimation. (c) and (d) Inspection of the diffraction character of identified regions of consistency for different labels (2 and 4) within the domain map.}
    \label{bcm-eval}
\end{figure}

This consistency map provides a visual summary for how invariant the different regions of the sample are to a change in the parameters of the workflow, an example for the dataset discussed in section \ref{HMA-sec} is shown in Figure \ref{bcm-eval}.

Pixels with values close to 1.0 consistently appear in clusters with the same sets of pixels, whereas pixels with low values tend to change which sets of pixels they are grouped with. Lower values in the consistency map correspond to regions which are sensitive to hyper-parameter choice.
%are change whether they are separated based on the granularity of the domain map. 
This allows the consistency map to be used to help assess if a selected domain map provides an appropriate segmentation for the desired experiment. For a selected domain map, each individual domain can be applied as a mask to the consistency map. Assessment of the values of the consistency map within one domain can be used to identify which regions (if any) within this domain are likely to change with a change of clustering parameters. The average patterns associated with each of these identified regions can then be calculated and a visual assessment can be made if the differences between these patterns should, in fact, constitute separation into discrete domains. This can be repeated for all the domains in a given map in order to increase the confidence in this map as a suitable representation of the data and that the best choice has been made for the number of centroids used in the segmentation. 

 Figure \ref{bcm-eval} shows an example of applying this protocol to two classes within a domain map. The top left image shows a domain map which was produced using 10 centroids identified using NMF. The top right shows the Statistical Consistency Map generated by comparing the map using 10 centroids, to a series of maps using 12,14,16,18,20,22 and 24 centroids, respectively. The bottom left and bottom right images, show the regions of the Statistical Consistency Map which correspond to domains 2 and 4 in the 10 centroid map. These are the domain-masked consistency maps. Within each of these two domain-masked consistency maps, it can be seen that there are different consistency regions: regions with higher consistency values (darker green); and regions with lower consistency values (lighter green). Overlaid on to these domain masked consistency maps are the mean diffraction patterns averaged over each of the consistency regions. For domain 2, five consistency regions are identified, and while the bottom-left-most of these patterns is more noisy than the others, all patterns show good agreement in reflection location and this domain could be considered satisfactory. Conversely, while domain 4 (bottom right) also shows five regions, the mean patterns associated with these regions are less consistent in peak position and intensity, suggesting that the map in Figure \ref{bcm-eval} (a), is not performing a satisfactory segmentation, by classing all these pixels as within the same domain. If, upon inspection, all the classes within a map produce patterns for the different consistency regions akin to the domain 2 (bottom left) then this should improve the confidence in the domain map as suitable. If, upon inspection, some of the classes within a map produce patterns for the different consistency regions more like domain 4 (bottom right) then this should reduce the confidence in the domain map and might suggest investigating the use of different parameters for generating the mapping.

\section{Comparison with Other Techniques}
In this section, the outcome of the workflow presented in this paper is compared to some other more established approaches for processing 4D-STEM datasets. The techniques in question are a direct NMF of the sample data (without use of a VAE or the hull clustering technique) and a Feature Extraction based approach\cite{Bruefach2022} utilising  Py4DSTEM\cite{Savitzky2021} for much of the data processing. For the sake of simplicity, the techniques are compared using the same single dataset, the precipitate on thin Al foil, but the technique is applied to both the raw dataset and also the precipitate masked dataset to see how this pre-processing step affects performance. 

\subsection{Direct NMF}
\subsubsection{Method}
NMF is used on the dataset to extract 16 principal components (the same number used in the workflow example above) and the patterns are compared against these components using a root-mean-squared (RMS) comparison and a SSI comparison.  
\subsubsection{Results}
\begin{figure}[htbp]
\centering
\includegraphics[width=\linewidth]{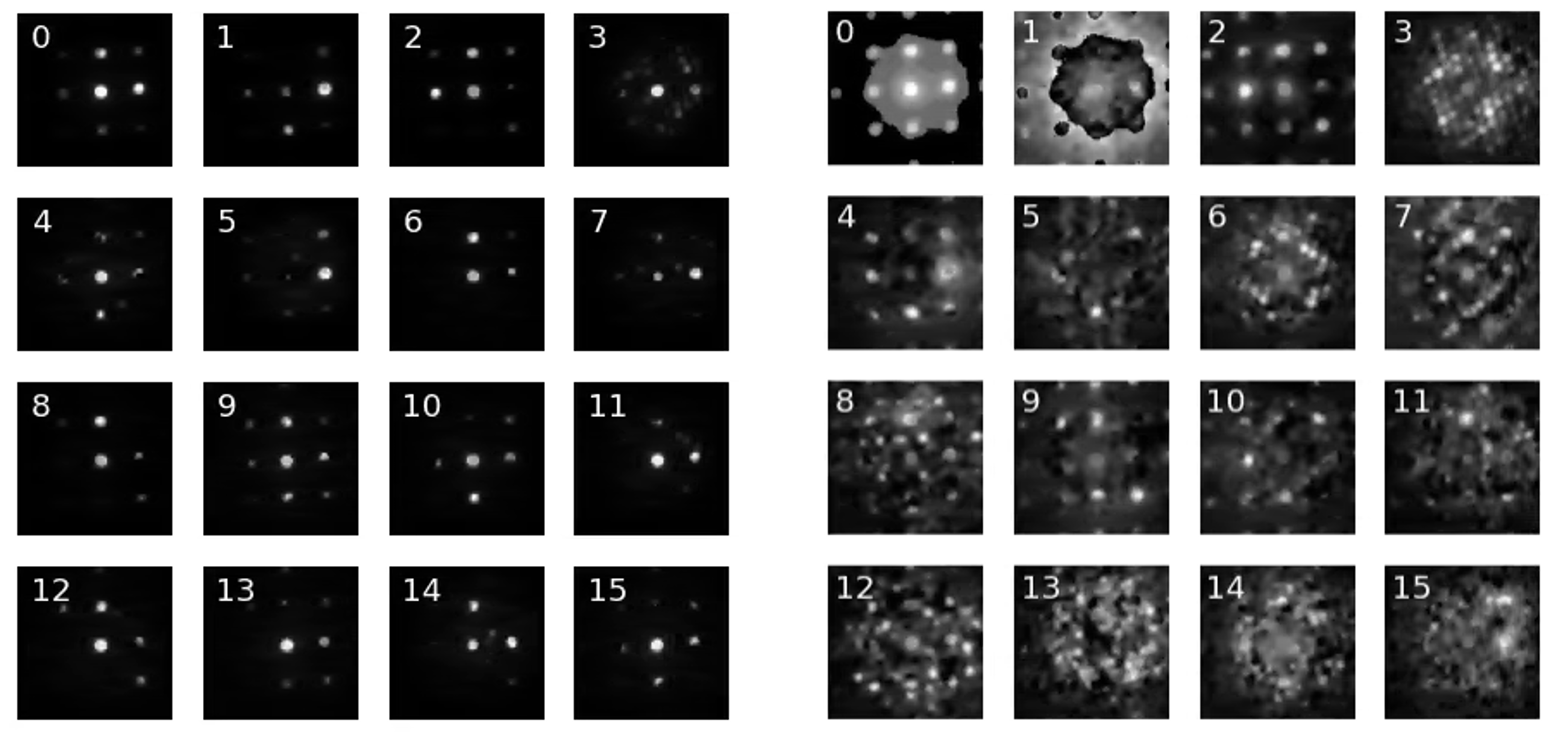}
\caption{The 16 principal components extracted: from the masked Precipitates on Al alloy dataset (left); and from the redistributed dataset generated from decoding the latent-space of the masked Precipitates on Al alloy dataset (right)}
\label{fig:pca-comp-factors}
\end{figure}

In Figure \ref{fig:pca-comp-factors}, the advantage of using the VAE to learn the sources of variance can be seen. As the VAE can be directed to learn to identify the precipitate peaks first, and can be truncated before learning the more subtle variations in the substrate matrix (as described in \ref{subsec:vae}), the rebalanced dataset generated will only show variance corresponding to the reflections of the precipitate. Thus, when performing NMF, the set of components produced will also be dominated by these precipitate peaks (Figure \ref{fig:pca-comp-factors}, right). This is not the case when performing NMF directly on the raw dataset, even if only using patterns from within the precipitate regions to perform the NMF, the resultant components are dominated by variation in the intensity of the substrate reflections (Figure \ref{fig:pca-comp-factors}, left).

\begin{figure}[htbp]
\centering
\includegraphics[width=\linewidth]{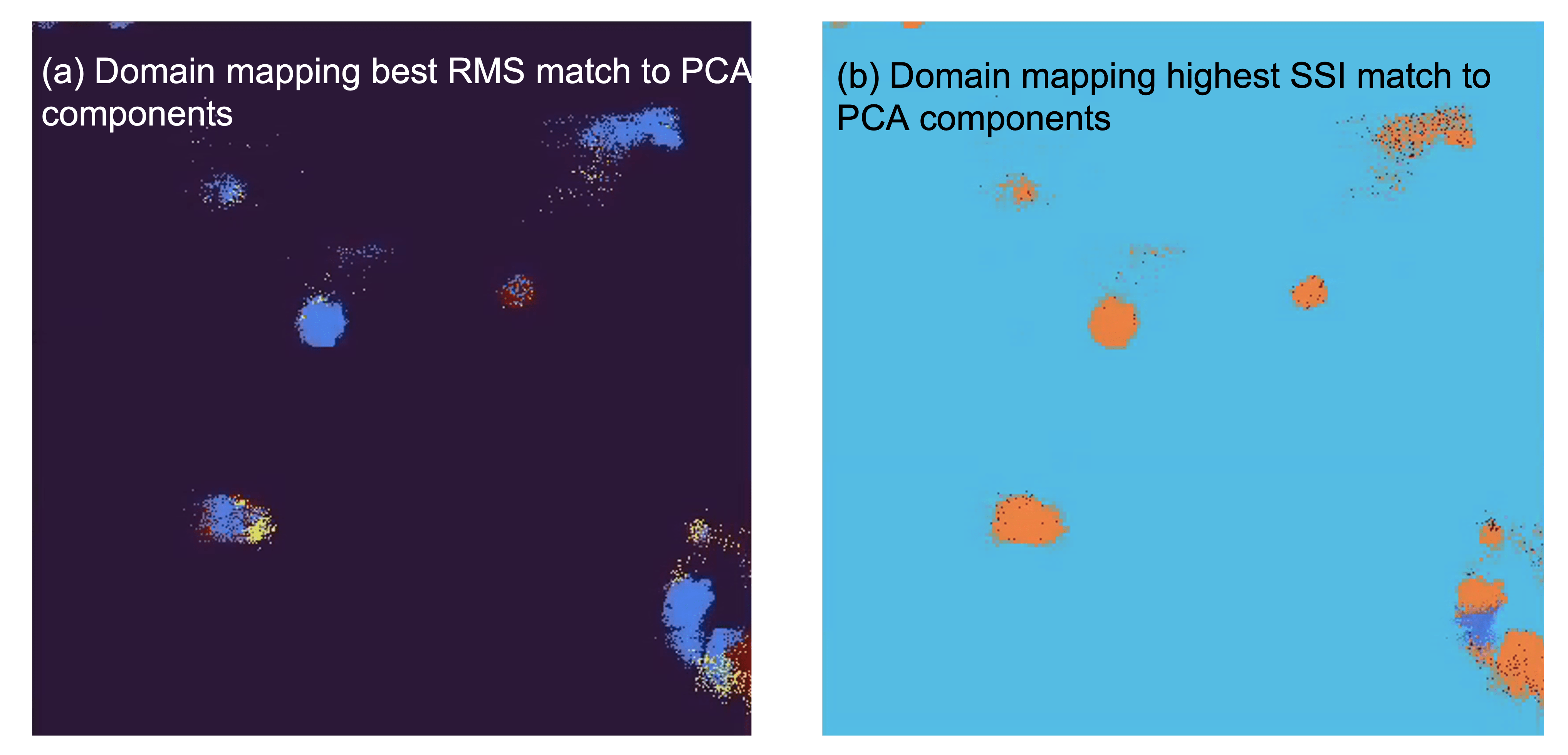}
\caption{The domain maps produced by direct NMF using a RMS comparison metric (left) and an SSI comparison metric (right).}
\label{fig:pca-results}
\end{figure}

In order to use the components to then generate clusters, a typical approach would be to compare the sample patterns to the principal components using RMS. Using SSI can provide a superior comparison metric but takes longer, and is still ultimately limited by the weak component detection and in this case produces an even poorer domain map than RMS (see Figure \ref{fig:pca-results}. Neither of the maps in Figure \ref{fig:pca-results} are able to differentiate between the different precipitates in a satisfactory manner. This can be seen explicitly by looking at the patterns in Figure \ref{fig:hmafit} (e) and (f). For example, the single patterns of pixels 3 and 12 are clearly from different orientations and the mean patterns in (f) corroborate this. Both methods in  Figure \ref{fig:pca-results} would have these pixels in the same domain (and would in-fact combine almost all of the domains identified in Figure \ref{fig:hmafit} into one).

\begin{figure}[htbp]
\centering
\includegraphics[width=\linewidth]{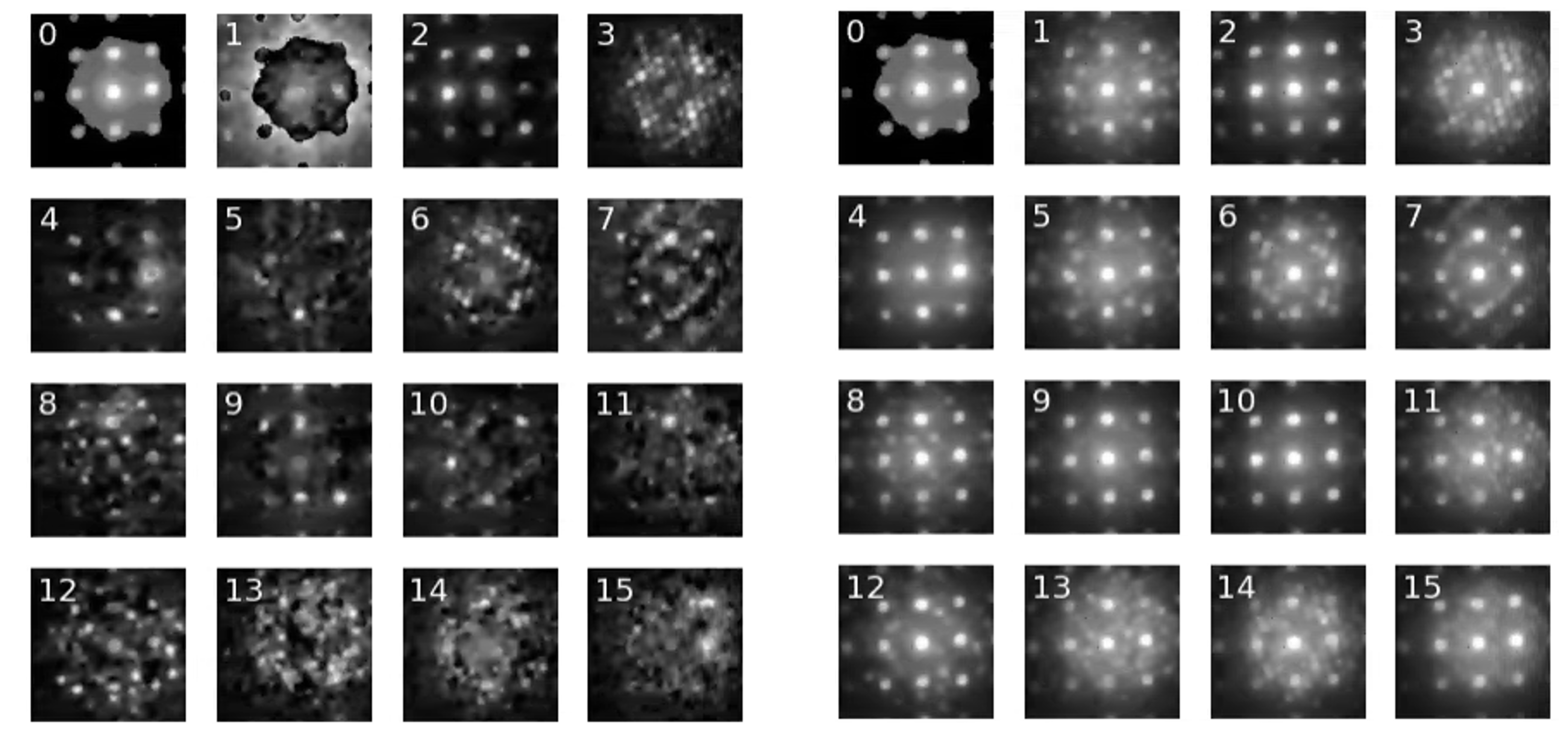}
\caption{The 16 principal components extracted from the redistributed dataset from the masked Precipitates on Al alloy latent-space and the centroid pattern used to represent each component}
\label{fig:pca2centroid}
\end{figure}

Inspecting the components generated during the centroid estimation, the benefit of the VAE workflow approach becomes apparent, as while the different components identified are indeed driven by the different precipitate patterns, the artifacts of the NMF limit the direct applicability of comparison metrics. Figure \ref{fig:pca2centroid} shows the centroid estimation of each of the components identified by the workflow. The use of these centroid patterns in the comparison metrics as opposed to the components identified by direct NMF in Figure \ref{fig:pca-comp-factors} accounts for the majority of the performance improvements. 

\subsection{Feature Extraction Approach}
\subsubsection{Method Overview}
An alternative contemporary approach is to use the workflow outlined in Bruefach et. al.\cite{Bruefach2022} This approach performs three methods of feature extraction on the dataset to represent the diffraction information and then for each, an iterative NMF decomposition\cite{Allen2021} to produce domain maps. The feature extraction methods are Radial Variance, Angular Averaging and Bragg Disk Detection. More information of the specifics are available in the manuscript. As before, the workflow is applied to both the raw thin foil Al alloy and the precipitate masked thin foil Al alloy datasets.

For the radial variance feature extraction, a polar elliptical transformation is applied to convert the Cartesian coordinates of the detector grid into polar coordinates. The variance for each row of pixels (corresponding to a single radius from the centre or a radial ring in the original pattern) is then calculated. 

The angular averaging requires finding the local minima of the radial integral of the maximum diffraction pattern. This is used to approximate the location of the radial rings of reflections. The largest and smallest rings are used to define a reflection range and every 5 rows are averaged to return the angular average.

The Bragg disk detection is performed using built in methods of py4DSTEM\cite{Savitzky2021}. It involves convolving a probe kernel with each pattern and detecting the maxima of the cross correlation to identify diffraction peaks. These peaks are then rasterised into a grid with $3 \times 3$ bins to reduce the dimensions and this grid is vectorised to produce the features list.

More information of the specifics are available in the manuscript [REF]. As before, the workflow is applied to both the raw thin foil Al alloy and the precipitate masked thin foil Al alloy datasets.

\subsubsection{Results}
\begin{figure}[htbp]
\centering
\includegraphics[width=\linewidth]{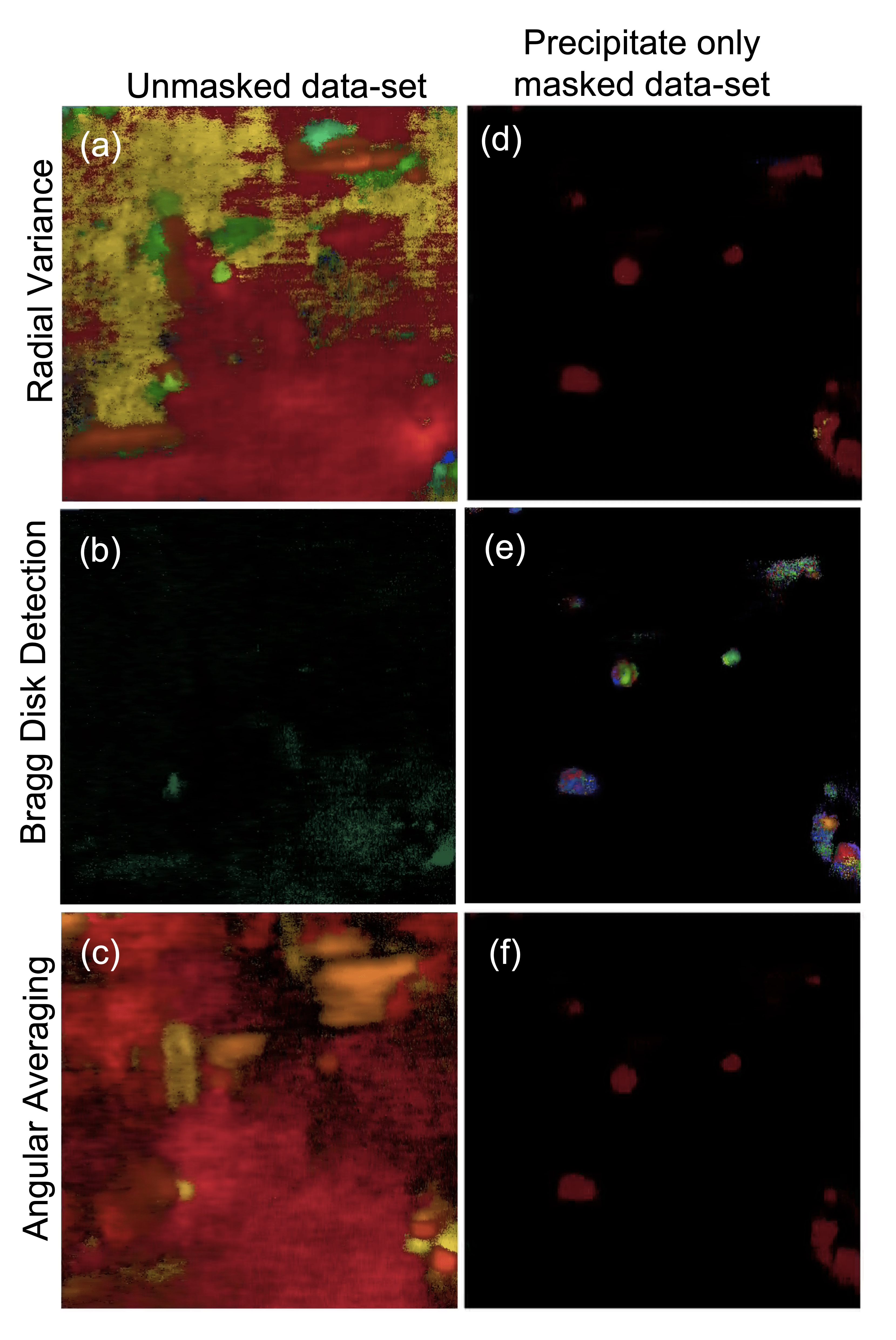}
\caption{The left column is the unmasked full thin foil Al alloy dataset clustered using (a) Radial Variance, (b) Bragg Disk Detection and (c) Angular Averaging. The right column is the precipitate only masked thin foil Al alloy dataset clustered using (a) Radial Variance, (b) Bragg Disk Detection and (c) Angular Averaging.}
\label{fig:HMA-alex-full}
\end{figure}

The results of the feature extraction workflow can be seen in Figure \ref{fig:HMA-alex-full}. The unmasked dataset shows that the substrate variation dominates the domain maps for all three feature approaches. For the masked dataset the Bragg Peak method performs the best, identifying some regions within precipitates, but the segmentations are noisy and the precipitates are over-segemented.  These shortcomings are likely due to the low precipitate peak intensity present in a single patterns resulting in these features being poorly expressed when principal components are identified using NMF. It is plausible that through optimal parameter selection the results of this method could be improved however, with the size of this dataset ($511 \times 511 \times 256 \times 256$) and the default processing pipeline, the iteration time for parameter tuning is prohibitive. An approximate comparison of the compute times, on the same hardware, for the Bragg Peak Feature Extraction workflow and our workflow are given in Table \ref{tab:py4dstem-time}. The run times were calculated running the code on a cluster on the Diamond Light Source jupyterhub\cite{jupyterhub} equipped with 16 CPU cores, 128\,GB RAM, Nvidia Tesla V100 GPU, Cuda 10.1. 

The iteration time is the time required to complete the workflow if a change is made to the given process. As completing the workflow is often required to evaluate the quality of the parameter change, this metric can be used to indicate how quickly parameter iterations can occur at each stage of the workflow. From the iteration times in Table \ref{tab:py4dstem-time}, it can be seen that parameter tuning, even for the last step of the Feature Extraction workflow, will take in the order of hours, whereas our workflow can be iterated from complete model retraining in under an hour with the clustering behaviour tuned in closer to 5 minutes per iteration. 

The memory usage for the two workflows was also calculated by running the same operations on the Diamond Light Source Hamilton Cluster equipped with Nvidia P100 GPU and 128\,GB RAM. The Bragg Peak Feature extraction used a maximum of 36\,GB of virtual memory (vmem) whereas the workflow in this paper used a maximum of 116\,GB of vmem.

\begin{table*}[t]
\centering
\resizebox{\textwidth}{!}{%
\begin{tabular}{lll|lll|}
\hline
\multicolumn{3}{|l|}{Bragg Peak Feature Extraction workflow} & \multicolumn{3}{l|}{VAE Workflow} \\ \hline
\multicolumn{1}{|l|}{Process} & \multicolumn{1}{l|}{Process Duration} & Iteration Time & \multicolumn{1}{l|}{Process} & \multicolumn{1}{l|}{Process Duration} & Iteration Time \\ \hline
\multicolumn{1}{|l|}{Bragg Disk Detection} & \multicolumn{1}{l|}{12hr 25min 9s} & 19hr 39min 31s & \multicolumn{1}{l|}{Model Training} & \multicolumn{1}{l|}{35min 33s} & 43min 16s \\
\multicolumn{1}{|l|}{Disk Intensity Calculation} & \multicolumn{1}{l|}{1hr 26min 54s} & 7hr 14min 22s & \multicolumn{1}{l|}{Pattern Pre-processing} & \multicolumn{1}{l|}{1min 46s} & 7min 43s \\
\multicolumn{1}{|l|}{Universal Thresholding} & \multicolumn{1}{l|}{1hr 26min 35s} & 5hr 47min 28s & \multicolumn{1}{l|}{Pattern Encoding} & \multicolumn{1}{l|}{32s} & 5min 57s \\
\multicolumn{1}{|l|}{Iterative NMF} & \multicolumn{1}{l|}{4hr 20min 53s} & 4hr 20min 53s & \multicolumn{1}{l|}{Boundary Point Estimation} & \multicolumn{1}{l|}{1min 14s} & 5min 25s \\ \cline{1-3}
 &  &  & \multicolumn{1}{l|}{Centroid Estimation} & \multicolumn{1}{l|}{1min 35s} & 4min 11s \\
 &  &  & \multicolumn{1}{l|}{Clustering} & \multicolumn{1}{l|}{2min 36s} & 2min 36s \\ \cline{4-6} 
\end{tabular}%
}
\caption{Comparison of approximate durations of processes in Bragg Peak Feature Extraction workflow with processes in the workflow present in this paper, on the same hardware. The iteration time given is the approximate time to complete the domain mapping if a parameter adjustment is desired for that process. The iteration time of the first process gives the approximate duration of the entire workflow.}
\label{tab:py4dstem-time}
\end{table*}

\section{Workflow Automation and Resource Requirements}
The discussions throughout the previous sections have intended to highlight the flexibility of the workflow in different use cases, but also emphasise that a versatile approach, requires directing towards producing domains optimised for the investigations of each experiment. This then promotes questions of how much intervention and resources are required for analysing each new dataset. The drive towards more autonomous workflows is an aim to promote higher throughput analysis and improve the statistical significance of the observations that can be drawn out of these domain maps. 

The first area for automation is at the VAE training. While curating the training data is required to optimally bias the encoder towards learning relevant variation, this processing pipeline should not require alteration from sample to sample within the same experiment. This means that once a pre-processing routine has been identified which satisfactorily biases the encoder, this routine can then be applied automatically to other datasets with the same desired domain separation. For example, for the dataset in Section \ref{p2-sec}, the pre-processing requires removal of the carbon support and vacuum patterns, and then balancing the dataset to increase the occurrence of high entropy patterns. With this identified, to produce the domain map of other LMO samples, the same pre-processing workflow could be applied automatically to each dataset to produce the training set. 

With the training of the VAE requiring the bulk of the total processing time, automating this process provides a significant improvement in the rate at which domain maps can be produced (see Iteration Time in Table \ref{tab:py4dstem-time}). The testing of different parameterisations in the clustering step can also be automated, to produce a set of domain maps which can be used for producing a Statistical Consistency Map. The final step of selecting whether an outputted domain map is suitable, however, currently requires human assessment.
The resource requirements of this process are high, due to the training of the VAE requiring GPU acceleration to execute in a workable time, and this is certainly a limitation of the workflow. To mitigate this training time, if there is access to sufficient memory, it is possible to aggregate all the pre-processed training datasets together and train one model on all of the data; this model should have a larger latent dimension to handle the added variance and avoid overly compressing the data. As a result of the architecture of the model, transfer learning can then be used to help improve the training time. As the model is made up of 2D Convolution (Conv2D) layers, followed by densely connected (Dense) layers, it is expected that the Conv2D layers learn the low level concepts and then these are passed to the Dense layers which learn to apply these concepts to the specific sample case. Therefore we can apply transfer learning to take the weights of the outer Conv2D layers of this bulk-trained model and apply them to the standard VAE architecture, which is then trained on the specific training dataset in reduced time. 

\section{Conclusions}

We have presented an automated workflow for analysing SEND datasets. The approach that we present while automated, is flexible enough to allow the adaptation of inductive biases in the model, to account for different experimental targets. We have demonstrated the power of our methodology first on a simulated dataset where ground truth is available for evaluation, then by extracting useful domain maps from a heterogeneous set of experiments where the data and the desired outcomes are significantly different. Our workflow was able to extract these maps with minimal human intervention (aside from deciding the aim of the experiment) and in a short time. We compared our workflow to other common approaches for SEND data analysis and showed that our approach is more versatile and more efficient,  ultimately yielding better more meaningful domain maps. Finally we also demonstrated a confidence map associated with our SEND analysis, that would allow the experimenter to identify and investigate regions where domain classifications are ambiguous. 

With the growth in the speed of data collection, automation of SEND data analysis is imperative. However, automation cannot come at the cost of expert biases of experimenters being able to drive the analysis. The workflow that we present balances automation against the ability to flexibly alter physical biases to drive the analysis. We believe that this synergistic interaction of humans and automated analysis will be critical for unlocking the power of SEND experiments to link information on the atomic and bulk scales in materials. 

\section*{Data Availability}
The datasets and model weights for both the Layered Metal Oxide (P2 Sample) and the Simulated Dataset are available at \url{https://doi.org/10.5281/zenodo.6839129}. The dataset and model weights for the Precipitates in Al alloy thin foil may be made available by our industrial collaborator upon reasonable request. Worked jupyter notebooks are available on Github\cite{coderepo}.

\section{Methods}
\label{sec:data}
The experimental datasets presented were collected at the Electron Physical Sciences Imaging Centre (ePSIC) at Diamond Light Source. The exact sample compositions and synthesis procedures are not included due to confidentiality requirements of our collaborators, but these are simply example SEND datasets and this information is not necessary for the workflow's application.

\subsection{Simulated dataset - Data Collection}
\label{subsec:simdatadetails}
The simulated dataset used a Crystallographic Information File (CIF) from the Inorganic Crystal Structure Database (ICSD)\cite{Allen1998} for \ce{NaCoO2} with space group $R-3m$\cite{Takahasi2003}. This CIF is used in conjunction with PyPrismatic's multislice algorithm\cite{DaCosta2021}\cite{Ophus2017}\cite{Pryor2017} to simulate patterns for the different zone axes using an accelerating voltage of 200\,kV, a sample thickness of 12 unit cells (approximately 10\,nm) and a convergence semi-angle of 2\,mrad. These patterns were then used to populate the a real space based on the ground truth map in Figure \ref{fig:simdatafull} (f). The bar was populated by using a weighted combination of the simulated pattern and the vacuum pattern, weighted by the horizontal coordinate of the pixel (the lower the coordinate, the higher the proportion of unscattered pattern). Region 6 was populated using a sampling of amorphous diffraction data from a real dataset. All patterns then had a Poissonian noise filter applied to simulate real detector noise. 

\subsection{Simulated dataset - VAE Training}
VAE training was performed on the Diamond Light Source jupyterhub\cite{jupyterhub} equipped with 16 CPU cores, 128\,GB RAM, Nvidia Tesla V100 GPU and Cuda 10.1. Training was performed on the full dataset of 87320, $128 \times 128$ diffraction patterns. As training was conducted on the full dataset performance was evaluated on the full dataset as the model was not required to perform on patterns outside of this domain. The training was performed using the Adam optimiser\cite{Adamoptim}, using a learning rate of 0.0001. The training was executed for 1000 epochs, with the model weights saved at each epoch that showed a reduction in validation loss.

\subsection{Precipitates on Al alloy foil - Data Collection}
This dataset was collect on the JEOL Grand-ARM 300F microscope at ePSIC, using pencil-beam alignment, operating at 200\,kV accelerating voltage. The 4D-STEM data was collected on a single chip MerlinEM detector from Quantum Detectors ($256 \times 256$ pixel array). The pencil beam was set up by turning the probe spherical aberration corrector optics off and reducing the probe convergence semi-angle by gradually changing the lens excitations in the pre-specimen optics to lower the cross-over of the beam. The final convergence semi-angle of the probe using the 10\,$\mu$m condenser lens aperture was around 1.5\,mrad.  

\subsection{Precipitates on Al alloy foil - VAE Training}
VAE training was performed on the Diamond Light Source jupyterhub\cite{jupyterhub} equipped with 16 CPU cores, 128\,GB RAM, Nvidia Tesla V100 GPU and Cuda 10.1. Training was performed on the a dataset of 8650, $128 \times 128$ diffraction patterns. The raw $256 \times 256$ patterns were cropped to select the central $128 \times 128$ pixels. The training dataset was sampled from the precipitate regions identified by the masking method described in \ref{HMA-sec}. The training was performed using the Adam optimiser\cite{Adamoptim}, using a learning rate of 0.0001. The training was executed for 1000 epochs, with the model weights saved at each epoch that showed a reduction in validation loss. 

\subsection{Layered Metal Oxide - Data Collection}
This dataset was collect on the JEOL Grand-ARM 300F at ePSIC, in similar optical arrangement as above, operated at 300\,kV accelerating voltage. A quad MerlinEM detector from Quantum Detectors was used $515 \times 515$. The probe convergence semi-angle using the 10\,$\mu$m condenser lens aperture was around 1.5\,mrad.

\subsection{Layered Metal Oxide - VAE Training}
VAE training was performed on the Diamond Light Source jupyterhub\cite{jupyterhub} equipped with 16 CPU cores, 128\,GB RAM, Nvidia Tesla V100 GPU and Cuda 10.1. Training was performed on the a dataset of 50000, $128 \times 128$ diffraction patterns. The raw $476 \times 476$ patterns were cropped to select the central $256 \times 256$ pixels and these patterns were downsampled by a factor of 2 to obtain $128 \times 128$ patterns. The training dataset was sampled using the Shannon Entropy of the patterns within the full dataset to define a probability distribution. Training performance was evaluated on a validation set containing 5000 randomly selected patterns. The training was performed using the Adam optimiser\cite{Adamoptim}, using a learning rate of 0.0001. The training was executed for 1000 epochs, with the model weights saved at each epoch that showed a reduction in validation loss.

\subsection{VAE Model Selection}
The final weight selection is done by visually inspecting, at various stages of the training, the decoded points from different regions in the latent-space. To aid this inspection, a regular grid of points, spanning the region of the latent-space populated by encoded data, can be sampled and passed through the decoder to form a new 4D-STEM 'latent-space dataset'. This new dataset then allows quick investigation of the patterns associated with the different regions of the latent-space. 
To select an appropriate model, the final set of weights from the training was selected, the data encoded and the distribution of the points in the latent-space viewed. Separated regions were observed in the encoded points and the patterns associated with these regions were inspected using the latent-space dataset. If the differences in the patterns associated with these regions were deemed to be experimentally relevant (showing meaningful variation in intensity and/or Bragg peak location) then the model was used. If the differences in the regions were not relevant (such as showing variation in background noise or insignificant variation in a single peak intensities) then the model was deemed to be overtrained and an earlier model was investigated. The selection of a model is a balance of time and performance. Some level of overfitting is managable due to the workflow's flexibility in domains selection, and for a preliminary high-throughput investigation, saving time on optimising the model performance is likely preferable. For an in-depth study of a sample, such as in this paper however, selecting a model that is well tuned to the experimental goals will give a better domain mapping performance.    

\section*{Acknowledgements}

We thank Diamond Light Source for access and support in use of the electron Physical Science Imaging Centre (Instrument E02 and proposal numbers EM19064 and MG28749) that contributed to the results presented here. AB gratefully acknowledges support from the joint Oxford-Diamond-STFC studentship scheme. TJW and WIFD also acknowledge the Faraday Institution (Grant No. FIRG018) for their funding contributions.

\section*{Author Contributions}
Based on the CRediT taxonomy the authors have made the following contributions:
AB - Conceptualization, Data curation, Formal Analysis, Investigation, Methodology, Software, Visualization, Writing – original draft, Writing – review \& editing \\
WIFD - Conceptualization, Funding acquisition, Project administration, Resources, Supervision, Writing – review \& editing
TJW - Conceptualization, Funding acquisition, Project administration, Resources, Supervision, Writing – review \& editing
MD - Conceptualization, Funding acquisition, Formal Analysis, Methodology, Project administration, Resources, Supervision, Writing – review \& editing
KTB - Conceptualization, Funding acquisition, Formal Analysis, Methodology, Project administration, Supervision, Writing – review \& editing

\section*{Conflicting Interests}
The authors have no conflicts of interest to declare that are relevant to the content of this article.

\appendix 
\section{Mathematical Operations}
\label{matop}
\subsection{Root Mean Squared Error}
The root mean squared error (RMSE) is given by Equation \ref{eq:rms}
\begin{equation}
\label{eq:rms}
    RMSE=\sqrt{\sum _{i}^{N}\frac{(\hat{x_i}-x_i)^2}{N}}
\end{equation}

where $\hat{x_i}$ is the expected value of x for i, $x_i$ is the observed value of x for i and $n$ is the number of observations.

\subsection{Kullback-Leibler Divergence}
Kullback-Leibler Divergence\cite{Kullback1951} is a statistical measure of the divergence between two distributions $P$ and $Q$ given by Equation \ref{eq:kl}

\begin{equation}
\label{eq:kl}
    D_{KL}(P\|Q)=\sum_iP(i)\text{ln}\left(\frac{P(i)}{Q(i)}\right)
\end{equation}

\subsection{Kernel Density Estimation}
\label{sec:kde}
Kernel Density Estimation (KDE) is a non-parameteric approach to estimating the population density from a sample. This is done using sklearn\cite{Brucher2011}, which utilises Tree searches for efficient Nearest Neighbour queries.

For the set of d-dimensional samples $x_{i}; i = 1, 2, ... , n$  in $\mathbb{R}^{d}$, sampled from a parent population with density function $p(x)$, the kernel density estimate at point $y$, $p_{k}(y)$ can be given by Equation \ref{eq:kde}

\begin{equation}
\label{eq:kde}
    p_{k}(y)= \sum _{i}^{N}K(y-x_i;h)
\end{equation}

where $K(x;h)\propto e^{-\frac{x^2}{2h^2}}$ for a Gaussian Kernel.

\subsection{Structural Similarity Index}
The Structural Similarity Index (SSI) is an image comparison metric\cite{Wang2009}\cite{Wang2004} implemented in skimage\cite{Walt2014}.

\subsection{Non-negative Matrix Factorisation}
Non-negative matrix factorisation is a means of factorising a matrix into a linear combination of a set of basis vectors such that 

\begin{equation}
\label{eq:nmf}
    V\approx WH
\end{equation}

where $V$ is the matrix, $W$ is the transformed data and $H$ is the basis vector. The constraint of NMF is that $V$,$W$ and $H$ are all entirely $>0$. Equation \ref{eq:nmf} does not have a closed form solution so can be solved by iteration using different algorithms. The algorithm used in this workflow is built in to PyXem\cite{Cautaerts2022}.

\nocite{*}

\bibliography{main}% Produces the bibliography via BibTeX.

\begin{thebibliography}{51}%
\makeatletter
\providecommand \@ifxundefined [1]{%
 \@ifx{#1\undefined}
}%
\providecommand \@ifnum [1]{%
 \ifnum #1\expandafter \@firstoftwo
 \else \expandafter \@secondoftwo
 \fi
}%
\providecommand \@ifx [1]{%
 \ifx #1\expandafter \@firstoftwo
 \else \expandafter \@secondoftwo
 \fi
}%
\providecommand \natexlab [1]{#1}%
\providecommand \enquote  [1]{``#1''}%
\providecommand \bibnamefont  [1]{#1}%
\providecommand \bibfnamefont [1]{#1}%
\providecommand \citenamefont [1]{#1}%
\providecommand \href@noop [0]{\@secondoftwo}%
\providecommand \href [0]{\begingroup \@sanitize@url \@href}%
\providecommand \@href[1]{\@@startlink{#1}\@@href}%
\providecommand \@@href[1]{\endgroup#1\@@endlink}%
\providecommand \@sanitize@url [0]{\catcode `\\12\catcode `\$12\catcode
  `\&12\catcode `\#12\catcode `\^12\catcode `\_12\catcode `\%12\relax}%
\providecommand \@@startlink[1]{}%
\providecommand \@@endlink[0]{}%
\providecommand \url  [0]{\begingroup\@sanitize@url \@url }%
\providecommand \@url [1]{\endgroup\@href {#1}{\urlprefix }}%
\providecommand \urlprefix  [0]{URL }%
\providecommand \Eprint [0]{\href }%
\providecommand \doibase [0]{https://doi.org/}%
\providecommand \selectlanguage [0]{\@gobble}%
\providecommand \bibinfo  [0]{\@secondoftwo}%
\providecommand \bibfield  [0]{\@secondoftwo}%
\providecommand \translation [1]{[#1]}%
\providecommand \BibitemOpen [0]{}%
\providecommand \bibitemStop [0]{}%
\providecommand \bibitemNoStop [0]{.\EOS\space}%
\providecommand \EOS [0]{\spacefactor3000\relax}%
\providecommand \BibitemShut  [1]{\csname bibitem#1\endcsname}%
\let\auto@bib@innerbib\@empty
%</preamble>
\bibitem [{\citenamefont {Ophus}\ \emph {et~al.}(2014)\citenamefont {Ophus},
  \citenamefont {Ercius}, \citenamefont {Sarahan}, \citenamefont {Czarnik},\
  and\ \citenamefont {Ciston}}]{Ophus2014}%
  \BibitemOpen
  \bibfield  {author} {\bibinfo {author} {\bibfnamefont {C.}~\bibnamefont
  {Ophus}}, \bibinfo {author} {\bibfnamefont {P.}~\bibnamefont {Ercius}},
  \bibinfo {author} {\bibfnamefont {M.}~\bibnamefont {Sarahan}}, \bibinfo
  {author} {\bibfnamefont {C.}~\bibnamefont {Czarnik}},\ and\ \bibinfo {author}
  {\bibfnamefont {J.}~\bibnamefont {Ciston}},\ }\bibfield  {title} {\bibinfo
  {title} {{Recording and using 4D-STEM datasets in materials science}},\
  }\href {https://doi.org/10.1017/S1431927614002037} {\bibfield  {journal}
  {\bibinfo  {journal} {Microsc. Microanal.}\ }\textbf {\bibinfo {volume}
  {20}},\ \bibinfo {pages} {62} (\bibinfo {year} {2014})}\BibitemShut {NoStop}%
\bibitem [{\citenamefont {Bunaciu}\ \emph {et~al.}(2015)\citenamefont
  {Bunaciu}, \citenamefont {gabriela Udriştioiu},\ and\ \citenamefont
  {Aboul-Enein}}]{Bunaciu2015}%
  \BibitemOpen
  \bibfield  {author} {\bibinfo {author} {\bibfnamefont {A.~A.}\ \bibnamefont
  {Bunaciu}}, \bibinfo {author} {\bibfnamefont {E.}~\bibnamefont {gabriela
  Udriştioiu}},\ and\ \bibinfo {author} {\bibfnamefont {H.~Y.}\ \bibnamefont
  {Aboul-Enein}},\ }\bibfield  {title} {\bibinfo {title} {{X-Ray Diffraction:
  Instrumentation and Applications}},\ }\href
  {https://doi.org/10.1080/10408347.2014.949616} {\bibfield  {journal}
  {\bibinfo  {journal} {Crit. Rev. Anal. Chem.}\ }\textbf {\bibinfo {volume}
  {45}},\ \bibinfo {pages} {289} (\bibinfo {year} {2015})}\BibitemShut
  {NoStop}%
\bibitem [{\citenamefont {Cheetham}\ and\ \citenamefont
  {Taylor}(1977)}]{Cheetam1977}%
  \BibitemOpen
  \bibfield  {author} {\bibinfo {author} {\bibfnamefont {A.~K.}\ \bibnamefont
  {Cheetham}}\ and\ \bibinfo {author} {\bibfnamefont {J.~C.}\ \bibnamefont
  {Taylor}},\ }\bibfield  {title} {\bibinfo {title} {{Profile analysis of
  powder neutron diffraction data: Its scope, limitations, and applications in
  solid state chemistry}},\ }\href
  {https://doi.org/10.1016/0022-4596(77)90203-1} {\bibfield  {journal}
  {\bibinfo  {journal} {J. Solid State Chem.}\ }\textbf {\bibinfo {volume}
  {21}},\ \bibinfo {pages} {253} (\bibinfo {year} {1977})}\BibitemShut
  {NoStop}%
\bibitem [{\citenamefont {James}\ and\ \citenamefont
  {Browning}(1999)}]{James1999}%
  \BibitemOpen
  \bibfield  {author} {\bibinfo {author} {\bibfnamefont {E.~M.}\ \bibnamefont
  {James}}\ and\ \bibinfo {author} {\bibfnamefont {N.~D.}\ \bibnamefont
  {Browning}},\ }\bibfield  {title} {\bibinfo {title} {{Practical aspects of
  atomic resolution imaging and analysis in STEM}},\ }\href
  {https://doi.org/10.1016/S0304-3991(99)00018-2} {\bibfield  {journal}
  {\bibinfo  {journal} {Ultramicroscopy}\ }\textbf {\bibinfo {volume} {78}},\
  \bibinfo {pages} {125} (\bibinfo {year} {1999})}\BibitemShut {NoStop}%
\bibitem [{\citenamefont {MacLaren}\ \emph {et~al.}(2020)\citenamefont
  {MacLaren}, \citenamefont {MacGregor}, \citenamefont {Allen},\ and\
  \citenamefont {Kirkland}}]{McLaren2020}%
  \BibitemOpen
  \bibfield  {author} {\bibinfo {author} {\bibfnamefont {I.}~\bibnamefont
  {MacLaren}}, \bibinfo {author} {\bibfnamefont {T.~A.}\ \bibnamefont
  {MacGregor}}, \bibinfo {author} {\bibfnamefont {C.~S.}\ \bibnamefont
  {Allen}},\ and\ \bibinfo {author} {\bibfnamefont {A.~I.}\ \bibnamefont
  {Kirkland}},\ }\bibfield  {title} {\bibinfo {title} {{Detectors-The ongoing
  revolution in scanning transmission electron microscopy and why this
  important to material characterization}},\ }\bibfield  {journal} {\bibinfo
  {journal} {APL Mater.}\ }\textbf {\bibinfo {volume} {8}},\ \href
  {https://doi.org/10.1063/5.0026992} {10.1063/5.0026992} (\bibinfo {year}
  {2020})\BibitemShut {NoStop}%
\bibitem [{\citenamefont {Ophus}(2019)}]{Ophus2019}%
  \BibitemOpen
  \bibfield  {author} {\bibinfo {author} {\bibfnamefont {C.}~\bibnamefont
  {Ophus}},\ }\bibfield  {title} {\bibinfo {title} {{Four-Dimensional Scanning
  Transmission Electron Microscopy (4D-STEM): From Scanning Nanodiffraction to
  Ptychography and Beyond}},\ }\href
  {https://doi.org/10.1017/S1431927619000497} {\bibfield  {journal} {\bibinfo
  {journal} {Microsc. Microanal.}\ ,\ \bibinfo {pages} {563}} (\bibinfo {year}
  {2019})}\BibitemShut {NoStop}%
\bibitem [{\citenamefont {Laulainen}\ \emph {et~al.}(2019)\citenamefont
  {Laulainen}, \citenamefont {Johnstone}, \citenamefont {Bogachev},
  \citenamefont {Collins}, \citenamefont {Longley}, \citenamefont {Bennett},\
  and\ \citenamefont {Midgley}}]{Laulainen2019}%
  \BibitemOpen
  \bibfield  {author} {\bibinfo {author} {\bibfnamefont {J.~E.~M.}\
  \bibnamefont {Laulainen}}, \bibinfo {author} {\bibfnamefont {D.~N.}\
  \bibnamefont {Johnstone}}, \bibinfo {author} {\bibfnamefont {I.}~\bibnamefont
  {Bogachev}}, \bibinfo {author} {\bibfnamefont {S.~M.}\ \bibnamefont
  {Collins}}, \bibinfo {author} {\bibfnamefont {L.}~\bibnamefont {Longley}},
  \bibinfo {author} {\bibfnamefont {T.~D.}\ \bibnamefont {Bennett}},\ and\
  \bibinfo {author} {\bibfnamefont {P.~A.}\ \bibnamefont {Midgley}},\
  }\bibfield  {title} {\bibinfo {title} {{Mapping Non-Crystalline Nanostructure
  in Beam Sensitive Systems With Low-dose Scanning Electron Pair Distribution
  Function Analysis}},\ }\href {https://doi.org/10.1017/s1431927619008912}
  {\bibfield  {journal} {\bibinfo  {journal} {Microsc. Microanal.}\ }\textbf
  {\bibinfo {volume} {25}},\ \bibinfo {pages} {1636} (\bibinfo {year}
  {2019})}\BibitemShut {NoStop}%
\bibitem [{\citenamefont {Johnstone}\ \emph {et~al.}(2020)\citenamefont
  {Johnstone}, \citenamefont {Firth}, \citenamefont {Grey}, \citenamefont
  {Midgley}, \citenamefont {Cliffe}, \citenamefont {Collins},\ and\
  \citenamefont {Johnstone}}]{Johnstone2020}%
  \BibitemOpen
  \bibfield  {author} {\bibinfo {author} {\bibfnamefont {D.~N.}\ \bibnamefont
  {Johnstone}}, \bibinfo {author} {\bibfnamefont {F.~C.}\ \bibnamefont
  {Firth}}, \bibinfo {author} {\bibfnamefont {C.~P.}\ \bibnamefont {Grey}},
  \bibinfo {author} {\bibfnamefont {P.~A.}\ \bibnamefont {Midgley}}, \bibinfo
  {author} {\bibfnamefont {M.~J.}\ \bibnamefont {Cliffe}}, \bibinfo {author}
  {\bibfnamefont {S.~M.}\ \bibnamefont {Collins}},\ and\ \bibinfo {author}
  {\bibfnamefont {D.~N.}\ \bibnamefont {Johnstone}},\ }\bibfield  {title}
  {\bibinfo {title} {{Direct Imaging of Correlated Defect Nanodomains in a
  Metal-Organic Framework}},\ }\href {https://doi.org/10.1021/jacs.0c04468}
  {\bibfield  {journal} {\bibinfo  {journal} {J. Am. Chem. Soc}\ }\textbf
  {\bibinfo {volume} {142}},\ \bibinfo {pages} {13081} (\bibinfo {year}
  {2020})}\BibitemShut {NoStop}%
\bibitem [{\citenamefont {Brunetti}\ \emph {et~al.}(2011)\citenamefont
  {Brunetti}, \citenamefont {Robert}, \citenamefont {Bayle-Guillemaud},
  \citenamefont {Rouvi{\`{e}}re}, \citenamefont {Rauch}, \citenamefont
  {Martin}, \citenamefont {Colin}, \citenamefont {Bertin},\ and\ \citenamefont
  {Cayron}}]{Brunetti2011}%
  \BibitemOpen
  \bibfield  {author} {\bibinfo {author} {\bibfnamefont {G.}~\bibnamefont
  {Brunetti}}, \bibinfo {author} {\bibfnamefont {D.}~\bibnamefont {Robert}},
  \bibinfo {author} {\bibfnamefont {P.}~\bibnamefont {Bayle-Guillemaud}},
  \bibinfo {author} {\bibfnamefont {J.~L.}\ \bibnamefont {Rouvi{\`{e}}re}},
  \bibinfo {author} {\bibfnamefont {E.~F.}\ \bibnamefont {Rauch}}, \bibinfo
  {author} {\bibfnamefont {J.~F.}\ \bibnamefont {Martin}}, \bibinfo {author}
  {\bibfnamefont {J.~F.}\ \bibnamefont {Colin}}, \bibinfo {author}
  {\bibfnamefont {F.}~\bibnamefont {Bertin}},\ and\ \bibinfo {author}
  {\bibfnamefont {C.}~\bibnamefont {Cayron}},\ }\bibfield  {title} {\bibinfo
  {title} {{Confirmation of the domino-cascade model by lifepo4/fepo 4
  precession electron diffraction}},\ }\href
  {https://doi.org/10.1021/cm201783z} {\bibfield  {journal} {\bibinfo
  {journal} {Chem. Mater.}\ }\textbf {\bibinfo {volume} {23}},\ \bibinfo
  {pages} {4515} (\bibinfo {year} {2011})}\BibitemShut {NoStop}%
\bibitem [{\citenamefont {Kobler}\ \emph {et~al.}(2013)\citenamefont {Kobler},
  \citenamefont {Kashiwar}, \citenamefont {Hahn},\ and\ \citenamefont
  {K{\"{u}}bel}}]{Kobler2013}%
  \BibitemOpen
  \bibfield  {author} {\bibinfo {author} {\bibfnamefont {A.}~\bibnamefont
  {Kobler}}, \bibinfo {author} {\bibfnamefont {A.}~\bibnamefont {Kashiwar}},
  \bibinfo {author} {\bibfnamefont {H.}~\bibnamefont {Hahn}},\ and\ \bibinfo
  {author} {\bibfnamefont {C.}~\bibnamefont {K{\"{u}}bel}},\ }\bibfield
  {title} {\bibinfo {title} {{Combination of in situ straining and ACOM TEM: A
  novel method for analysis of plastic deformation of nanocrystalline
  metals}},\ }\href {https://doi.org/10.1016/j.ultramic.2012.12.019} {\bibfield
   {journal} {\bibinfo  {journal} {Ultramicroscopy}\ }\textbf {\bibinfo
  {volume} {128}},\ \bibinfo {pages} {68} (\bibinfo {year} {2013})}\BibitemShut
  {NoStop}%
\bibitem [{\citenamefont {Gallagher-Jones}\ \emph {et~al.}(2019)\citenamefont
  {Gallagher-Jones}, \citenamefont {Ophus}, \citenamefont {Bustillo},
  \citenamefont {Boyer}, \citenamefont {Panova}, \citenamefont {Glynn},
  \citenamefont {Zee}, \citenamefont {Ciston}, \citenamefont {Mancia},
  \citenamefont {Minor},\ and\ \citenamefont {Rodriguez}}]{Gallagher2019}%
  \BibitemOpen
  \bibfield  {author} {\bibinfo {author} {\bibfnamefont {M.}~\bibnamefont
  {Gallagher-Jones}}, \bibinfo {author} {\bibfnamefont {C.}~\bibnamefont
  {Ophus}}, \bibinfo {author} {\bibfnamefont {K.~C.}\ \bibnamefont {Bustillo}},
  \bibinfo {author} {\bibfnamefont {D.~R.}\ \bibnamefont {Boyer}}, \bibinfo
  {author} {\bibfnamefont {O.}~\bibnamefont {Panova}}, \bibinfo {author}
  {\bibfnamefont {C.}~\bibnamefont {Glynn}}, \bibinfo {author} {\bibfnamefont
  {C.~T.}\ \bibnamefont {Zee}}, \bibinfo {author} {\bibfnamefont
  {J.}~\bibnamefont {Ciston}}, \bibinfo {author} {\bibfnamefont {K.~C.}\
  \bibnamefont {Mancia}}, \bibinfo {author} {\bibfnamefont {A.~M.}\
  \bibnamefont {Minor}},\ and\ \bibinfo {author} {\bibfnamefont {J.~A.}\
  \bibnamefont {Rodriguez}},\ }\bibfield  {title} {\bibinfo {title} {{Nanoscale
  mosaicity revealed in peptide microcrystals by scanning electron
  nanodiffraction}},\ }\href {https://doi.org/10.1038/s42003-018-0263-8}
  {\bibfield  {journal} {\bibinfo  {journal} {Commun. Biol.}\ }\textbf
  {\bibinfo {volume} {2}},\ \bibinfo {pages} {1} (\bibinfo {year}
  {2019})}\BibitemShut {NoStop}%
\bibitem [{\citenamefont {{Nalin Mehta}}\ \emph {et~al.}(2020)\citenamefont
  {{Nalin Mehta}}, \citenamefont {Gauquelin}, \citenamefont {Nord},
  \citenamefont {Orekhov}, \citenamefont {Bender}, \citenamefont {Cerbu},
  \citenamefont {Verbeeck},\ and\ \citenamefont {Vandervorst}}]{Nalin2020}%
  \BibitemOpen
  \bibfield  {author} {\bibinfo {author} {\bibfnamefont {A.}~\bibnamefont
  {{Nalin Mehta}}}, \bibinfo {author} {\bibfnamefont {N.}~\bibnamefont
  {Gauquelin}}, \bibinfo {author} {\bibfnamefont {M.}~\bibnamefont {Nord}},
  \bibinfo {author} {\bibfnamefont {A.}~\bibnamefont {Orekhov}}, \bibinfo
  {author} {\bibfnamefont {H.}~\bibnamefont {Bender}}, \bibinfo {author}
  {\bibfnamefont {D.}~\bibnamefont {Cerbu}}, \bibinfo {author} {\bibfnamefont
  {J.}~\bibnamefont {Verbeeck}},\ and\ \bibinfo {author} {\bibfnamefont
  {W.}~\bibnamefont {Vandervorst}},\ }\bibfield  {title} {\bibinfo {title}
  {{Unravelling stacking order in epitaxial bilayer MX2using 4D-STEM with
  unsupervised learning}},\ }\bibfield  {journal} {\bibinfo  {journal}
  {Nanotechnology}\ }\textbf {\bibinfo {volume} {31}},\ \href
  {https://doi.org/10.1088/1361-6528/aba5b6} {10.1088/1361-6528/aba5b6}
  (\bibinfo {year} {2020})\BibitemShut {NoStop}%
\bibitem [{\citenamefont {Ortiz}\ \emph {et~al.}(2020)\citenamefont {Ortiz},
  \citenamefont {Zhu}, \citenamefont {Dou},\ and\ \citenamefont
  {Hwang}}]{Ortiz2020}%
  \BibitemOpen
  \bibfield  {author} {\bibinfo {author} {\bibfnamefont {G.~C.}\ \bibnamefont
  {Ortiz}}, \bibinfo {author} {\bibfnamefont {M.}~\bibnamefont {Zhu}}, \bibinfo
  {author} {\bibfnamefont {L.}~\bibnamefont {Dou}},\ and\ \bibinfo {author}
  {\bibfnamefont {J.}~\bibnamefont {Hwang}},\ }\bibfield  {title} {\bibinfo
  {title} {{4D-STEM Quantification of Nanoscale Ordered Domains in Organic
  Semiconducting Polymers}},\ }\href
  {https://doi.org/10.1017/s1431927620019170} {\bibfield  {journal} {\bibinfo
  {journal} {Microsc. Microanal.}\ }\textbf {\bibinfo {volume} {26}},\ \bibinfo
  {pages} {1740} (\bibinfo {year} {2020})}\BibitemShut {NoStop}%
\bibitem [{\citenamefont {Kalinin}\ \emph {et~al.}(2021)\citenamefont
  {Kalinin}, \citenamefont {Oxley}, \citenamefont {Valleti}, \citenamefont
  {Zhang}, \citenamefont {Hermann}, \citenamefont {Zheng}, \citenamefont
  {Zhang}, \citenamefont {Eres}, \citenamefont {Vasudevan},\ and\ \citenamefont
  {Ziatdinov}}]{Khalinin2021}%
  \BibitemOpen
  \bibfield  {author} {\bibinfo {author} {\bibfnamefont {S.~V.}\ \bibnamefont
  {Kalinin}}, \bibinfo {author} {\bibfnamefont {M.~P.}\ \bibnamefont {Oxley}},
  \bibinfo {author} {\bibfnamefont {M.}~\bibnamefont {Valleti}}, \bibinfo
  {author} {\bibfnamefont {J.}~\bibnamefont {Zhang}}, \bibinfo {author}
  {\bibfnamefont {R.~P.}\ \bibnamefont {Hermann}}, \bibinfo {author}
  {\bibfnamefont {H.}~\bibnamefont {Zheng}}, \bibinfo {author} {\bibfnamefont
  {W.}~\bibnamefont {Zhang}}, \bibinfo {author} {\bibfnamefont
  {G.}~\bibnamefont {Eres}}, \bibinfo {author} {\bibfnamefont {R.~K.}\
  \bibnamefont {Vasudevan}},\ and\ \bibinfo {author} {\bibfnamefont
  {M.}~\bibnamefont {Ziatdinov}},\ }\bibfield  {title} {\bibinfo {title} {{Deep
  Bayesian local crystallography}},\ }\bibfield  {journal} {\bibinfo  {journal}
  {Npj Comput. Mater.}\ }\textbf {\bibinfo {volume} {7}},\ \href
  {https://doi.org/10.1038/s41524-021-00621-6} {10.1038/s41524-021-00621-6}
  (\bibinfo {year} {2021}),\ \Eprint {https://arxiv.org/abs/2012.07134}
  {arXiv:2012.07134} \BibitemShut {NoStop}%
\bibitem [{\citenamefont {Ophus}\ \emph {et~al.}(2022)\citenamefont {Ophus},
  \citenamefont {Zeltmann}, \citenamefont {Bruefach}, \citenamefont {Rakowski},
  \citenamefont {Savitzky}, \citenamefont {Minor},\ and\ \citenamefont
  {Scott}}]{Ophus2022}%
  \BibitemOpen
  \bibfield  {author} {\bibinfo {author} {\bibfnamefont {C.}~\bibnamefont
  {Ophus}}, \bibinfo {author} {\bibfnamefont {S.~E.}\ \bibnamefont {Zeltmann}},
  \bibinfo {author} {\bibfnamefont {A.}~\bibnamefont {Bruefach}}, \bibinfo
  {author} {\bibfnamefont {A.}~\bibnamefont {Rakowski}}, \bibinfo {author}
  {\bibfnamefont {B.~H.}\ \bibnamefont {Savitzky}}, \bibinfo {author}
  {\bibfnamefont {A.~M.}\ \bibnamefont {Minor}},\ and\ \bibinfo {author}
  {\bibfnamefont {M.~C.}\ \bibnamefont {Scott}},\ }\bibfield  {title} {\bibinfo
  {title} {{Automated Crystal Orientation Mapping in py4DSTEM using Sparse
  Correlation Matching}},\ }\href {https://doi.org/10.1017/S1431927622000101}
  {\bibfield  {journal} {\bibinfo  {journal} {Microsc. Microanal.}\ }\textbf
  {\bibinfo {volume} {28}},\ \bibinfo {pages} {390} (\bibinfo {year} {2022})},\
  \Eprint {https://arxiv.org/abs/2111.00171} {arXiv:2111.00171} \BibitemShut
  {NoStop}%
\bibitem [{\citenamefont {Cautaerts}\ \emph {et~al.}(2022)\citenamefont
  {Cautaerts}, \citenamefont {Crout}, \citenamefont {{\AA}nes}, \citenamefont
  {Prestat}, \citenamefont {Jeong}, \citenamefont {Dehm},\ and\ \citenamefont
  {Liebscher}}]{Cautaerts2022}%
  \BibitemOpen
  \bibfield  {author} {\bibinfo {author} {\bibfnamefont {N.}~\bibnamefont
  {Cautaerts}}, \bibinfo {author} {\bibfnamefont {P.}~\bibnamefont {Crout}},
  \bibinfo {author} {\bibfnamefont {H.~W.}\ \bibnamefont {{\AA}nes}}, \bibinfo
  {author} {\bibfnamefont {E.}~\bibnamefont {Prestat}}, \bibinfo {author}
  {\bibfnamefont {J.}~\bibnamefont {Jeong}}, \bibinfo {author} {\bibfnamefont
  {G.}~\bibnamefont {Dehm}},\ and\ \bibinfo {author} {\bibfnamefont {C.~H.}\
  \bibnamefont {Liebscher}},\ }\bibfield  {title} {\bibinfo {title} {{Free,
  flexible and fast: Orientation mapping using the multi-core and
  GPU-accelerated template matching capabilities in the Python-based open
  source 4D-STEM analysis toolbox Pyxem}},\ }\href
  {https://doi.org/10.1016/j.ultramic.2022.113517} {\bibfield  {journal}
  {\bibinfo  {journal} {Ultramicroscopy}\ }\textbf {\bibinfo {volume} {237}},\
  \bibinfo {pages} {113517} (\bibinfo {year} {2022})},\ \Eprint
  {https://arxiv.org/abs/2111.07347} {arXiv:2111.07347} \BibitemShut {NoStop}%
\bibitem [{\citenamefont {Meyer}\ \emph {et~al.}(2021)\citenamefont {Meyer},
  \citenamefont {Kressdorf}, \citenamefont {Roddatis}, \citenamefont
  {Hoffmann}, \citenamefont {Jooss},\ and\ \citenamefont {Seibt}}]{Meyer2021}%
  \BibitemOpen
  \bibfield  {author} {\bibinfo {author} {\bibfnamefont {T.}~\bibnamefont
  {Meyer}}, \bibinfo {author} {\bibfnamefont {B.}~\bibnamefont {Kressdorf}},
  \bibinfo {author} {\bibfnamefont {V.}~\bibnamefont {Roddatis}}, \bibinfo
  {author} {\bibfnamefont {J.}~\bibnamefont {Hoffmann}}, \bibinfo {author}
  {\bibfnamefont {C.}~\bibnamefont {Jooss}},\ and\ \bibinfo {author}
  {\bibfnamefont {M.}~\bibnamefont {Seibt}},\ }\bibfield  {title} {\bibinfo
  {title} {{Phase Transitions in a Perovskite Thin Film Studied by
  Environmental In Situ Heating Nano-Beam Electron Diffraction}},\ }\bibfield
  {journal} {\bibinfo  {journal} {Small Methods}\ }\textbf {\bibinfo {volume}
  {5}},\ \href {https://doi.org/10.1002/smtd.202100464}
  {10.1002/smtd.202100464} (\bibinfo {year} {2021}),\ \Eprint
  {https://arxiv.org/abs/2007.14882} {arXiv:2007.14882} \BibitemShut {NoStop}%
\bibitem [{\citenamefont {Liu}\ \emph {et~al.}(2020)\citenamefont {Liu},
  \citenamefont {Deng}, \citenamefont {Gao}, \citenamefont {Chen},
  \citenamefont {Yin}, \citenamefont {Yang}, \citenamefont {Zou}, \citenamefont
  {Hou},\ and\ \citenamefont {Ji}}]{Liu2020}%
  \BibitemOpen
  \bibfield  {author} {\bibinfo {author} {\bibfnamefont {H.}~\bibnamefont
  {Liu}}, \bibinfo {author} {\bibfnamefont {W.}~\bibnamefont {Deng}}, \bibinfo
  {author} {\bibfnamefont {X.}~\bibnamefont {Gao}}, \bibinfo {author}
  {\bibfnamefont {J.}~\bibnamefont {Chen}}, \bibinfo {author} {\bibfnamefont
  {S.}~\bibnamefont {Yin}}, \bibinfo {author} {\bibfnamefont {L.}~\bibnamefont
  {Yang}}, \bibinfo {author} {\bibfnamefont {G.}~\bibnamefont {Zou}}, \bibinfo
  {author} {\bibfnamefont {H.}~\bibnamefont {Hou}},\ and\ \bibinfo {author}
  {\bibfnamefont {X.}~\bibnamefont {Ji}},\ }\bibfield  {title} {\bibinfo
  {title} {{Manganese‐based layered oxide cathodes for sodium ion
  batteries}},\ }\href {https://doi.org/10.1002/nano.202000030} {\bibfield
  {journal} {\bibinfo  {journal} {Nano Select}\ }\textbf {\bibinfo {volume}
  {1}},\ \bibinfo {pages} {200} (\bibinfo {year} {2020})}\BibitemShut {NoStop}%
\bibitem [{\citenamefont {Valery}\ \emph {et~al.}(2017)\citenamefont {Valery},
  \citenamefont {Rauch}, \citenamefont {Cl{\'{E}}ment},\ and\ \citenamefont
  {Lorut}}]{Valery2017}%
  \BibitemOpen
  \bibfield  {author} {\bibinfo {author} {\bibfnamefont {A.}~\bibnamefont
  {Valery}}, \bibinfo {author} {\bibfnamefont {E.~F.}\ \bibnamefont {Rauch}},
  \bibinfo {author} {\bibfnamefont {L.}~\bibnamefont {Cl{\'{E}}ment}},\ and\
  \bibinfo {author} {\bibfnamefont {F.}~\bibnamefont {Lorut}},\ }\bibfield
  {title} {\bibinfo {title} {{Retrieving overlapping crystals information from
  TEM nano-beam electron diffraction patterns}},\ }\href
  {https://doi.org/10.1111/jmi.12599} {\bibfield  {journal} {\bibinfo
  {journal} {J. Microsc.}\ }\textbf {\bibinfo {volume} {268}},\ \bibinfo
  {pages} {208} (\bibinfo {year} {2017})}\BibitemShut {NoStop}%
\bibitem [{\citenamefont {Vincent}\ and\ \citenamefont
  {Midgley}(1994)}]{Vincent1994}%
  \BibitemOpen
  \bibfield  {author} {\bibinfo {author} {\bibfnamefont {R.}~\bibnamefont
  {Vincent}}\ and\ \bibinfo {author} {\bibfnamefont {P.~A.}\ \bibnamefont
  {Midgley}},\ }\bibfield  {title} {\bibinfo {title} {{Double conical
  beam-rocking system for measurement of integrated electron diffraction
  intensities}},\ }\href {https://doi.org/10.1016/0304-3991(94)90039-6}
  {\bibfield  {journal} {\bibinfo  {journal} {Ultramicroscopy}\ }\textbf
  {\bibinfo {volume} {53}},\ \bibinfo {pages} {271} (\bibinfo {year}
  {1994})}\BibitemShut {NoStop}%
\bibitem [{\citenamefont {Bruefach}\ \emph {et~al.}(2022)\citenamefont
  {Bruefach}, \citenamefont {Ophus},\ and\ \citenamefont
  {Scott}}]{Bruefach2022}%
  \BibitemOpen
  \bibfield  {author} {\bibinfo {author} {\bibfnamefont {A.}~\bibnamefont
  {Bruefach}}, \bibinfo {author} {\bibfnamefont {C.}~\bibnamefont {Ophus}},\
  and\ \bibinfo {author} {\bibfnamefont {M.~C.}\ \bibnamefont {Scott}},\
  }\bibfield  {title} {\bibinfo {title} {Analysis of interpretable data
  representations for 4d-stem using unsupervised learning},\ }in\ \href@noop {}
  {\emph {\bibinfo {booktitle} {Proceedings of Microscopy and Microanalysis,
  Portland, Oregon, USA, 31 July – 4 August, 2022}}}\ (\bibinfo {year}
  {2022})\BibitemShut {NoStop}%
\bibitem [{\citenamefont {Shi}\ \emph {et~al.}(2021)\citenamefont {Shi},
  \citenamefont {Cao}, \citenamefont {Rehn}, \citenamefont {Bae}, \citenamefont
  {Kim}, \citenamefont {Jones}, \citenamefont {Muller},\ and\ \citenamefont
  {Han}}]{Shi2022}%
  \BibitemOpen
  \bibfield  {author} {\bibinfo {author} {\bibfnamefont {C.}~\bibnamefont
  {Shi}}, \bibinfo {author} {\bibfnamefont {M.~C.}\ \bibnamefont {Cao}},
  \bibinfo {author} {\bibfnamefont {S.~M.}\ \bibnamefont {Rehn}}, \bibinfo
  {author} {\bibfnamefont {S.-H.}\ \bibnamefont {Bae}}, \bibinfo {author}
  {\bibfnamefont {J.}~\bibnamefont {Kim}}, \bibinfo {author} {\bibfnamefont
  {M.~R.}\ \bibnamefont {Jones}}, \bibinfo {author} {\bibfnamefont {D.~A.}\
  \bibnamefont {Muller}},\ and\ \bibinfo {author} {\bibfnamefont
  {Y.}~\bibnamefont {Han}},\ }\bibfield  {title} {\bibinfo {title} {{Uncovering
  Material Deformations via Machine Learning Combined with Four-Dimensional
  Scanning Transmission Electron Microscopy}},\ }\href
  {https://doi.org/10.1038/s41524-022-00793-9} {\bibfield  {journal} {\bibinfo
  {journal} {Npj Comput. Mater.}\ }\textbf {\bibinfo {volume} {8}},\ \bibinfo
  {pages} {114} (\bibinfo {year} {2021})},\ \Eprint
  {https://arxiv.org/abs/2111.06496} {arXiv:2111.06496} \BibitemShut {NoStop}%
\bibitem [{\citenamefont {Oviedo}\ \emph {et~al.}(2021)\citenamefont {Oviedo},
  \citenamefont {Ferres}, \citenamefont {Buonassisi},\ and\ \citenamefont
  {Butler}}]{oviedo2021interpretable}%
  \BibitemOpen
  \bibfield  {author} {\bibinfo {author} {\bibfnamefont {F.}~\bibnamefont
  {Oviedo}}, \bibinfo {author} {\bibfnamefont {J.~L.}\ \bibnamefont {Ferres}},
  \bibinfo {author} {\bibfnamefont {T.}~\bibnamefont {Buonassisi}},\ and\
  \bibinfo {author} {\bibfnamefont {K.~T.}\ \bibnamefont {Butler}},\ }\bibfield
   {title} {\bibinfo {title} {Interpretable and explainable machine learning
  for materials science and chemistry},\ }\href@noop {} {\bibfield  {journal}
  {\bibinfo  {journal} {Accounts of Materials Research}\ } (\bibinfo {year}
  {2021})}\BibitemShut {NoStop}%
\bibitem [{\citenamefont {Liu}\ \emph {et~al.}(2022)\citenamefont {Liu},
  \citenamefont {Kelley}, \citenamefont {Vasudevan}, \citenamefont {Funakubo},
  \citenamefont {Ziatdinov},\ and\ \citenamefont
  {Kalinin}}]{liu2022experimental}%
  \BibitemOpen
  \bibfield  {author} {\bibinfo {author} {\bibfnamefont {Y.}~\bibnamefont
  {Liu}}, \bibinfo {author} {\bibfnamefont {K.~P.}\ \bibnamefont {Kelley}},
  \bibinfo {author} {\bibfnamefont {R.~K.}\ \bibnamefont {Vasudevan}}, \bibinfo
  {author} {\bibfnamefont {H.}~\bibnamefont {Funakubo}}, \bibinfo {author}
  {\bibfnamefont {M.~A.}\ \bibnamefont {Ziatdinov}},\ and\ \bibinfo {author}
  {\bibfnamefont {S.~V.}\ \bibnamefont {Kalinin}},\ }\bibfield  {title}
  {\bibinfo {title} {Experimental discovery of structure--property
  relationships in ferroelectric materials via active learning},\ }\href@noop
  {} {\bibfield  {journal} {\bibinfo  {journal} {Nature Machine Intelligence}\
  }\textbf {\bibinfo {volume} {4}},\ \bibinfo {pages} {341} (\bibinfo {year}
  {2022})}\BibitemShut {NoStop}%
\bibitem [{\citenamefont {Allotey}\ \emph {et~al.}(2021)\citenamefont
  {Allotey}, \citenamefont {Butler},\ and\ \citenamefont
  {Thiyagalingam}}]{allotey2021entropy}%
  \BibitemOpen
  \bibfield  {author} {\bibinfo {author} {\bibfnamefont {J.}~\bibnamefont
  {Allotey}}, \bibinfo {author} {\bibfnamefont {K.~T.}\ \bibnamefont
  {Butler}},\ and\ \bibinfo {author} {\bibfnamefont {J.}~\bibnamefont
  {Thiyagalingam}},\ }\bibfield  {title} {\bibinfo {title} {Entropy-based
  active learning of graph neural network surrogate models for materials
  properties},\ }\href@noop {} {\bibfield  {journal} {\bibinfo  {journal} {The
  Journal of Chemical Physics}\ }\textbf {\bibinfo {volume} {155}},\ \bibinfo
  {pages} {174116} (\bibinfo {year} {2021})}\BibitemShut {NoStop}%
\bibitem [{\citenamefont {Bridger}(2022)}]{coderepo}%
  \BibitemOpen
  \bibfield  {author} {\bibinfo {author} {\bibfnamefont {A.}~\bibnamefont
  {Bridger}},\ }\href@noop {} {\bibinfo {title} {Stemsegment}},\ \bibinfo
  {howpublished} {\url{https://github.com/andy-bridger/stemsegment}} (\bibinfo
  {year} {2022})\BibitemShut {NoStop}%
\bibitem [{\citenamefont {Wang}\ and\ \citenamefont {Bovik}(2009)}]{Wang2009}%
  \BibitemOpen
  \bibfield  {author} {\bibinfo {author} {\bibfnamefont {Z.}~\bibnamefont
  {Wang}}\ and\ \bibinfo {author} {\bibfnamefont {A.~C.}\ \bibnamefont
  {Bovik}},\ }\bibfield  {title} {\bibinfo {title} {{Mean Squared Error : Love
  It or Leave It ?}},\ }\href@noop {} {\bibfield  {journal} {\bibinfo
  {journal} {IEEE Signal Process. Mag.}\ ,\ \bibinfo {pages} {98}} (\bibinfo
  {year} {2009})}\BibitemShut {NoStop}%
\bibitem [{\citenamefont {Wang}\ \emph {et~al.}(2004)\citenamefont {Wang},
  \citenamefont {Bovik}, \citenamefont {Sheikh},\ and\ \citenamefont
  {Simoncelli}}]{Wang2004}%
  \BibitemOpen
  \bibfield  {author} {\bibinfo {author} {\bibfnamefont {Z.}~\bibnamefont
  {Wang}}, \bibinfo {author} {\bibfnamefont {A.~C.}\ \bibnamefont {Bovik}},
  \bibinfo {author} {\bibfnamefont {H.~R.}\ \bibnamefont {Sheikh}},\ and\
  \bibinfo {author} {\bibfnamefont {E.~P.}\ \bibnamefont {Simoncelli}},\
  }\bibfield  {title} {\bibinfo {title} {{Image quality assessment: From error
  visibility to structural similarity}},\ }\href
  {https://doi.org/10.1109/TIP.2003.819861} {\bibfield  {journal} {\bibinfo
  {journal} {IEEE Trans. Image Process.}\ }\textbf {\bibinfo {volume} {13}},\
  \bibinfo {pages} {600} (\bibinfo {year} {2004})}\BibitemShut {NoStop}%
\bibitem [{\citenamefont {Gallinari}\ \emph {et~al.}(1987)\citenamefont
  {Gallinari}, \citenamefont {Lecun}, \citenamefont {Thiria},\ and\
  \citenamefont {{Fogelman Soulie}}}]{Gallinari1987}%
  \BibitemOpen
  \bibfield  {author} {\bibinfo {author} {\bibfnamefont {P.}~\bibnamefont
  {Gallinari}}, \bibinfo {author} {\bibfnamefont {Y.}~\bibnamefont {Lecun}},
  \bibinfo {author} {\bibfnamefont {S.}~\bibnamefont {Thiria}},\ and\ \bibinfo
  {author} {\bibfnamefont {F.}~\bibnamefont {{Fogelman Soulie}}},\ }\bibfield
  {title} {\bibinfo {title} {Memoires associatives distribuees: Une comparaison
  (distributed associative memories: A comparison)},\ }in\ \href@noop {} {\emph
  {\bibinfo {booktitle} {Proceedings of COGNITIVA 87, Paris, La Villette, May
  1987}}}\ (\bibinfo  {publisher} {Cesta-Afcet},\ \bibinfo {year}
  {1987})\BibitemShut {NoStop}%
\bibitem [{\citenamefont {Lecun}(1987)}]{Lecun1987}%
  \BibitemOpen
  \bibfield  {author} {\bibinfo {author} {\bibfnamefont {Y.}~\bibnamefont
  {Lecun}},\ }\href@noop {} {\emph {\bibinfo {title} {PhD thesis: Modeles
  connexionnistes de l'apprentissage (connectionist learning models)}}}\
  (\bibinfo  {publisher} {Universite P. et M. Curie (Paris 6)},\ \bibinfo
  {year} {1987})\BibitemShut {NoStop}%
\bibitem [{\citenamefont {Rumelhart}\ \emph {et~al.}(1986)\citenamefont
  {Rumelhart}, \citenamefont {Hinton},\ and\ \citenamefont
  {Williams}}]{Rumelhart1986}%
  \BibitemOpen
  \bibfield  {author} {\bibinfo {author} {\bibfnamefont {D.~E.}\ \bibnamefont
  {Rumelhart}}, \bibinfo {author} {\bibfnamefont {G.~E.}\ \bibnamefont
  {Hinton}},\ and\ \bibinfo {author} {\bibfnamefont {R.~J.}\ \bibnamefont
  {Williams}},\ }\bibinfo {title} {Learning internal representations by error
  propagation},\ in\ \href@noop {} {\emph {\bibinfo {booktitle} {Parallel
  Distributed Processing: Explorations in the Microstructure of Cognition, Vol.
  1: Foundations}}}\ (\bibinfo  {publisher} {MIT Press},\ \bibinfo {address}
  {Cambridge, MA, USA},\ \bibinfo {year} {1986})\ p.\ \bibinfo {pages}
  {318–362}\BibitemShut {NoStop}%
\bibitem [{\citenamefont {Dayan}\ \emph {et~al.}(1995)\citenamefont {Dayan},
  \citenamefont {Hinton}, \citenamefont {Neal},\ and\ \citenamefont
  {Zemel}}]{Dayan1995}%
  \BibitemOpen
  \bibfield  {author} {\bibinfo {author} {\bibfnamefont {P.}~\bibnamefont
  {Dayan}}, \bibinfo {author} {\bibfnamefont {G.~E.}\ \bibnamefont {Hinton}},
  \bibinfo {author} {\bibfnamefont {R.~M.}\ \bibnamefont {Neal}},\ and\
  \bibinfo {author} {\bibfnamefont {R.~S.}\ \bibnamefont {Zemel}},\ }\bibfield
  {title} {\bibinfo {title} {{The Helmholtz machine.}},\ }\href
  {https://doi.org/10.1162/neco.1995.7.5.889} {\bibfield  {journal} {\bibinfo
  {journal} {Neural Comput.}\ }\textbf {\bibinfo {volume} {7}},\ \bibinfo
  {pages} {889} (\bibinfo {year} {1995})}\BibitemShut {NoStop}%
\bibitem [{\citenamefont {Abadi}\ \emph {et~al.}(2015)\citenamefont {Abadi},
  \citenamefont {Agarwal}, \citenamefont {Barham}, \citenamefont {Brevdo},
  \citenamefont {Chen}, \citenamefont {Citro}, \citenamefont {Corrado},
  \citenamefont {Davis}, \citenamefont {Dean}, \citenamefont {Devin},
  \citenamefont {Ghemawat}, \citenamefont {Goodfellow}, \citenamefont {Harp},
  \citenamefont {Irving}, \citenamefont {Isard}, \citenamefont {Jia},
  \citenamefont {Jozefowicz}, \citenamefont {Kaiser}, \citenamefont {Kudlur},
  \citenamefont {Levenberg}, \citenamefont {Man\'{e}}, \citenamefont {Monga},
  \citenamefont {Moore}, \citenamefont {Murray}, \citenamefont {Olah},
  \citenamefont {Schuster}, \citenamefont {Shlens}, \citenamefont {Steiner},
  \citenamefont {Sutskever}, \citenamefont {Talwar}, \citenamefont {Tucker},
  \citenamefont {Vanhoucke}, \citenamefont {Vasudevan}, \citenamefont
  {Vi\'{e}gas}, \citenamefont {Vinyals}, \citenamefont {Warden}, \citenamefont
  {Wattenberg}, \citenamefont {Wicke}, \citenamefont {Yu},\ and\ \citenamefont
  {Zheng}}]{Tensorflow}%
  \BibitemOpen
  \bibfield  {author} {\bibinfo {author} {\bibfnamefont {M.}~\bibnamefont
  {Abadi}}, \bibinfo {author} {\bibfnamefont {A.}~\bibnamefont {Agarwal}},
  \bibinfo {author} {\bibfnamefont {P.}~\bibnamefont {Barham}}, \bibinfo
  {author} {\bibfnamefont {E.}~\bibnamefont {Brevdo}}, \bibinfo {author}
  {\bibfnamefont {Z.}~\bibnamefont {Chen}}, \bibinfo {author} {\bibfnamefont
  {C.}~\bibnamefont {Citro}}, \bibinfo {author} {\bibfnamefont {G.~S.}\
  \bibnamefont {Corrado}}, \bibinfo {author} {\bibfnamefont {A.}~\bibnamefont
  {Davis}}, \bibinfo {author} {\bibfnamefont {J.}~\bibnamefont {Dean}},
  \bibinfo {author} {\bibfnamefont {M.}~\bibnamefont {Devin}}, \bibinfo
  {author} {\bibfnamefont {S.}~\bibnamefont {Ghemawat}}, \bibinfo {author}
  {\bibfnamefont {I.}~\bibnamefont {Goodfellow}}, \bibinfo {author}
  {\bibfnamefont {A.}~\bibnamefont {Harp}}, \bibinfo {author} {\bibfnamefont
  {G.}~\bibnamefont {Irving}}, \bibinfo {author} {\bibfnamefont
  {M.}~\bibnamefont {Isard}}, \bibinfo {author} {\bibfnamefont
  {Y.}~\bibnamefont {Jia}}, \bibinfo {author} {\bibfnamefont {R.}~\bibnamefont
  {Jozefowicz}}, \bibinfo {author} {\bibfnamefont {L.}~\bibnamefont {Kaiser}},
  \bibinfo {author} {\bibfnamefont {M.}~\bibnamefont {Kudlur}}, \bibinfo
  {author} {\bibfnamefont {J.}~\bibnamefont {Levenberg}}, \bibinfo {author}
  {\bibfnamefont {D.}~\bibnamefont {Man\'{e}}}, \bibinfo {author}
  {\bibfnamefont {R.}~\bibnamefont {Monga}}, \bibinfo {author} {\bibfnamefont
  {S.}~\bibnamefont {Moore}}, \bibinfo {author} {\bibfnamefont
  {D.}~\bibnamefont {Murray}}, \bibinfo {author} {\bibfnamefont
  {C.}~\bibnamefont {Olah}}, \bibinfo {author} {\bibfnamefont {M.}~\bibnamefont
  {Schuster}}, \bibinfo {author} {\bibfnamefont {J.}~\bibnamefont {Shlens}},
  \bibinfo {author} {\bibfnamefont {B.}~\bibnamefont {Steiner}}, \bibinfo
  {author} {\bibfnamefont {I.}~\bibnamefont {Sutskever}}, \bibinfo {author}
  {\bibfnamefont {K.}~\bibnamefont {Talwar}}, \bibinfo {author} {\bibfnamefont
  {P.}~\bibnamefont {Tucker}}, \bibinfo {author} {\bibfnamefont
  {V.}~\bibnamefont {Vanhoucke}}, \bibinfo {author} {\bibfnamefont
  {V.}~\bibnamefont {Vasudevan}}, \bibinfo {author} {\bibfnamefont
  {F.}~\bibnamefont {Vi\'{e}gas}}, \bibinfo {author} {\bibfnamefont
  {O.}~\bibnamefont {Vinyals}}, \bibinfo {author} {\bibfnamefont
  {P.}~\bibnamefont {Warden}}, \bibinfo {author} {\bibfnamefont
  {M.}~\bibnamefont {Wattenberg}}, \bibinfo {author} {\bibfnamefont
  {M.}~\bibnamefont {Wicke}}, \bibinfo {author} {\bibfnamefont
  {Y.}~\bibnamefont {Yu}},\ and\ \bibinfo {author} {\bibfnamefont
  {X.}~\bibnamefont {Zheng}},\ }\href {https://www.tensorflow.org/} {\bibinfo
  {title} {{TensorFlow}: Large-scale machine learning on heterogeneous
  systems}} (\bibinfo {year} {2015}),\ \bibinfo {note} {software available from
  tensorflow.org}\BibitemShut {NoStop}%
\bibitem [{\citenamefont {Kullback}\ and\ \citenamefont
  {Leibler}(1951)}]{Kullback1951}%
  \BibitemOpen
  \bibfield  {author} {\bibinfo {author} {\bibfnamefont {S.}~\bibnamefont
  {Kullback}}\ and\ \bibinfo {author} {\bibfnamefont {R.~A.}\ \bibnamefont
  {Leibler}},\ }\bibfield  {title} {\bibinfo {title} {{On Information and
  Sufficiency}},\ }\href@noop {} {\bibfield  {journal} {\bibinfo  {journal}
  {Ann. Math. Stat.}\ }\textbf {\bibinfo {volume} {22}},\ \bibinfo {pages} {79}
  (\bibinfo {year} {1951})}\BibitemShut {NoStop}%
\bibitem [{\citenamefont {Brucher}\ \emph {et~al.}(2011)\citenamefont
  {Brucher}, \citenamefont {Perrot},\ and\ \citenamefont
  {Duchesnay}}]{Brucher2011}%
  \BibitemOpen
  \bibfield  {author} {\bibinfo {author} {\bibfnamefont {M.}~\bibnamefont
  {Brucher}}, \bibinfo {author} {\bibfnamefont {M.}~\bibnamefont {Perrot}},\
  and\ \bibinfo {author} {\bibfnamefont {E.}~\bibnamefont {Duchesnay}},\
  }\bibfield  {title} {\bibinfo {title} {{Scikit-learn: Machine Learning in
  Python}},\ }\href {https://doi.org/10.1289/EHP4713} {\bibfield  {journal}
  {\bibinfo  {journal} {J. Mach. Learn Res.}\ }\textbf {\bibinfo {volume}
  {12}},\ \bibinfo {pages} {2825} (\bibinfo {year} {2011})}\BibitemShut
  {NoStop}%
\bibitem [{\citenamefont {Hien}\ \emph {et~al.}(2015)\citenamefont {Hien},
  \citenamefont {Tuan}, \citenamefont {At},\ and\ \citenamefont
  {Son}}]{Hien2015}%
  \BibitemOpen
  \bibfield  {author} {\bibinfo {author} {\bibfnamefont {T.~D.}\ \bibnamefont
  {Hien}}, \bibinfo {author} {\bibfnamefont {D.~V.}\ \bibnamefont {Tuan}},
  \bibinfo {author} {\bibfnamefont {P.~V.}\ \bibnamefont {At}},\ and\ \bibinfo
  {author} {\bibfnamefont {L.~H.}\ \bibnamefont {Son}},\ }\bibfield  {title}
  {\bibinfo {title} {{Novel algorithm for non-negative matrix factorization}},\
  }\href {https://doi.org/10.1142/S1793005715400013} {\bibfield  {journal}
  {\bibinfo  {journal} {New Math. Nat. Comput.}\ }\textbf {\bibinfo {volume}
  {11}},\ \bibinfo {pages} {121} (\bibinfo {year} {2015})}\BibitemShut
  {NoStop}%
\bibitem [{\citenamefont {Sunde}(2020)}]{Sunde2020}%
  \BibitemOpen
  \bibfield  {author} {\bibinfo {author} {\bibfnamefont {J.~K.}\ \bibnamefont
  {Sunde}},\ }\emph {\bibinfo {title} {{The Effect of Elevated Temperatures on
  Precipitation in Aluminium Alloys}}},\ \href@noop {} {\bibinfo {type} {Thesis
  for the degree of philosophiae doctor}},\ \bibinfo  {school} {Norwegian
  University of Science and Technology} (\bibinfo {year} {2020})\BibitemShut
  {NoStop}%
\bibitem [{\citenamefont {Shannon}(1948)}]{Shannon1948}%
  \BibitemOpen
  \bibfield  {author} {\bibinfo {author} {\bibfnamefont {C.~E.}\ \bibnamefont
  {Shannon}},\ }\bibfield  {title} {\bibinfo {title} {{A Mathematical Theory of
  Communication}},\ }\href {https://doi.org/10.1002/j.1538-7305.1948.tb00917.x}
  {\bibfield  {journal} {\bibinfo  {journal} {Bell Syst. tech. j.}\ }\textbf
  {\bibinfo {volume} {27}},\ \bibinfo {pages} {623} (\bibinfo {year}
  {1948})}\BibitemShut {NoStop}%
\bibitem [{\citenamefont {Slouf}\ \emph {et~al.}(2021)\citenamefont {Slouf},
  \citenamefont {Skoupy}, \citenamefont {Pavlova},\ and\ \citenamefont
  {Krzyzanek}}]{Slouf2021}%
  \BibitemOpen
  \bibfield  {author} {\bibinfo {author} {\bibfnamefont {M.}~\bibnamefont
  {Slouf}}, \bibinfo {author} {\bibfnamefont {R.}~\bibnamefont {Skoupy}},
  \bibinfo {author} {\bibfnamefont {E.}~\bibnamefont {Pavlova}},\ and\ \bibinfo
  {author} {\bibfnamefont {V.}~\bibnamefont {Krzyzanek}},\ }\bibfield  {title}
  {\bibinfo {title} {{High resolution powder electron diffraction in scanning
  electron microscopy}},\ }\bibfield  {journal} {\bibinfo  {journal}
  {Materials}\ }\textbf {\bibinfo {volume} {14}},\ \href
  {https://doi.org/10.3390/ma14247550} {10.3390/ma14247550} (\bibinfo {year}
  {2021})\BibitemShut {NoStop}%
\bibitem [{\citenamefont {Savitzky}\ \emph {et~al.}(2021)\citenamefont
  {Savitzky}, \citenamefont {Zeltmann}, \citenamefont {Hughes}, \citenamefont
  {Brown}, \citenamefont {Zhao}, \citenamefont {Pelz}, \citenamefont {Pekin},
  \citenamefont {Barnard}, \citenamefont {Donohue}, \citenamefont {{Rangel
  Dacosta}}, \citenamefont {Kennedy}, \citenamefont {Xie}, \citenamefont
  {Janish}, \citenamefont {Schneider}, \citenamefont {Herring}, \citenamefont
  {Gopal}, \citenamefont {Anapolsky}, \citenamefont {Dhall}, \citenamefont
  {Bustillo}, \citenamefont {Ercius}, \citenamefont {Scott}, \citenamefont
  {Ciston}, \citenamefont {Minor},\ and\ \citenamefont {Ophus}}]{Savitzky2021}%
  \BibitemOpen
  \bibfield  {author} {\bibinfo {author} {\bibfnamefont {B.~H.}\ \bibnamefont
  {Savitzky}}, \bibinfo {author} {\bibfnamefont {S.~E.}\ \bibnamefont
  {Zeltmann}}, \bibinfo {author} {\bibfnamefont {L.~A.}\ \bibnamefont
  {Hughes}}, \bibinfo {author} {\bibfnamefont {H.~G.}\ \bibnamefont {Brown}},
  \bibinfo {author} {\bibfnamefont {S.}~\bibnamefont {Zhao}}, \bibinfo {author}
  {\bibfnamefont {P.~M.}\ \bibnamefont {Pelz}}, \bibinfo {author}
  {\bibfnamefont {T.~C.}\ \bibnamefont {Pekin}}, \bibinfo {author}
  {\bibfnamefont {E.~S.}\ \bibnamefont {Barnard}}, \bibinfo {author}
  {\bibfnamefont {J.}~\bibnamefont {Donohue}}, \bibinfo {author} {\bibfnamefont
  {L.}~\bibnamefont {{Rangel Dacosta}}}, \bibinfo {author} {\bibfnamefont
  {E.}~\bibnamefont {Kennedy}}, \bibinfo {author} {\bibfnamefont
  {Y.}~\bibnamefont {Xie}}, \bibinfo {author} {\bibfnamefont {M.~T.}\
  \bibnamefont {Janish}}, \bibinfo {author} {\bibfnamefont {M.~M.}\
  \bibnamefont {Schneider}}, \bibinfo {author} {\bibfnamefont {P.}~\bibnamefont
  {Herring}}, \bibinfo {author} {\bibfnamefont {C.}~\bibnamefont {Gopal}},
  \bibinfo {author} {\bibfnamefont {A.}~\bibnamefont {Anapolsky}}, \bibinfo
  {author} {\bibfnamefont {R.}~\bibnamefont {Dhall}}, \bibinfo {author}
  {\bibfnamefont {K.~C.}\ \bibnamefont {Bustillo}}, \bibinfo {author}
  {\bibfnamefont {P.}~\bibnamefont {Ercius}}, \bibinfo {author} {\bibfnamefont
  {M.~C.}\ \bibnamefont {Scott}}, \bibinfo {author} {\bibfnamefont
  {J.}~\bibnamefont {Ciston}}, \bibinfo {author} {\bibfnamefont {A.~M.}\
  \bibnamefont {Minor}},\ and\ \bibinfo {author} {\bibfnamefont
  {C.}~\bibnamefont {Ophus}},\ }\bibfield  {title} {\bibinfo {title}
  {{Py4DSTEM: A Software Package for Four-Dimensional Scanning Transmission
  Electron Microscopy Data Analysis}},\ }\href
  {https://doi.org/10.1017/S1431927621000477} {\bibfield  {journal} {\bibinfo
  {journal} {Microscopy and Microanalysis}\ }\textbf {\bibinfo {volume} {27}},\
  \bibinfo {pages} {712} (\bibinfo {year} {2021})}\BibitemShut {NoStop}%
\bibitem [{\citenamefont {Allen}\ \emph {et~al.}(2021)\citenamefont {Allen},
  \citenamefont {Pekin}, \citenamefont {Persaud}, \citenamefont {Rozeveld},
  \citenamefont {Meyers}, \citenamefont {Ciston}, \citenamefont {Ophus},\ and\
  \citenamefont {Minor}}]{Allen2021}%
  \BibitemOpen
  \bibfield  {author} {\bibinfo {author} {\bibfnamefont {F.~I.}\ \bibnamefont
  {Allen}}, \bibinfo {author} {\bibfnamefont {T.~C.}\ \bibnamefont {Pekin}},
  \bibinfo {author} {\bibfnamefont {A.}~\bibnamefont {Persaud}}, \bibinfo
  {author} {\bibfnamefont {S.~J.}\ \bibnamefont {Rozeveld}}, \bibinfo {author}
  {\bibfnamefont {G.~F.}\ \bibnamefont {Meyers}}, \bibinfo {author}
  {\bibfnamefont {J.}~\bibnamefont {Ciston}}, \bibinfo {author} {\bibfnamefont
  {C.}~\bibnamefont {Ophus}},\ and\ \bibinfo {author} {\bibfnamefont {A.~M.}\
  \bibnamefont {Minor}},\ }\bibfield  {title} {\bibinfo {title} {{Fast Grain
  Mapping with Sub-Nanometer Resolution Using 4D-STEM with Grain Classification
  by Principal Component Analysis and Non-Negative Matrix Factorization}},\
  }\href {https://doi.org/10.1017/S1431927621011946} {\bibfield  {journal}
  {\bibinfo  {journal} {Microsc. Microanal.}\ }\textbf {\bibinfo {volume}
  {27}},\ \bibinfo {pages} {794} (\bibinfo {year} {2021})},\ \Eprint
  {https://arxiv.org/abs/2103.07076} {arXiv:2103.07076} \BibitemShut {NoStop}%
\bibitem [{jup(2022)}]{jupyterhub}%
  \BibitemOpen
  \href {https://www.jupyterhub.diamond.ac.uk/} {\bibinfo {title} {Diamond
  light source jupyterhub}} (\bibinfo {year} {2022})\BibitemShut {NoStop}%
\bibitem [{\citenamefont {Allen}(1998)}]{Allen1998}%
  \BibitemOpen
  \bibfield  {author} {\bibinfo {author} {\bibfnamefont {F.~H.}\ \bibnamefont
  {Allen}},\ }\bibfield  {title} {\bibinfo {title} {{The Development, Status
  and Scientific Impact of Crystallographic Databases}},\ }\href
  {https://doi.org/10.1107/S0108767398010563} {\bibfield  {journal} {\bibinfo
  {journal} {Acta Crystallogr. A}\ }\textbf {\bibinfo {volume} {54}},\ \bibinfo
  {pages} {758} (\bibinfo {year} {1998})}\BibitemShut {NoStop}%
\bibitem [{\citenamefont {Takahashi}\ \emph {et~al.}(2003)\citenamefont
  {Takahashi}, \citenamefont {Gotoh},\ and\ \citenamefont
  {Akimoto}}]{Takahasi2003}%
  \BibitemOpen
  \bibfield  {author} {\bibinfo {author} {\bibfnamefont {Y.}~\bibnamefont
  {Takahashi}}, \bibinfo {author} {\bibfnamefont {Y.}~\bibnamefont {Gotoh}},\
  and\ \bibinfo {author} {\bibfnamefont {J.}~\bibnamefont {Akimoto}},\
  }\bibfield  {title} {\bibinfo {title} {{Single-crystal growth, crystal and
  electronic structure of NaCoO2}},\ }\href
  {https://doi.org/10.1016/S0022-4596(02)00042-7} {\bibfield  {journal}
  {\bibinfo  {journal} {J. Solid State Chem.}\ }\textbf {\bibinfo {volume}
  {172}},\ \bibinfo {pages} {22} (\bibinfo {year} {2003})}\BibitemShut
  {NoStop}%
\bibitem [{\citenamefont {{Rangel DaCosta}}\ \emph {et~al.}(2021)\citenamefont
  {{Rangel DaCosta}}, \citenamefont {Brown}, \citenamefont {Pelz},
  \citenamefont {Rakowski}, \citenamefont {Barber}, \citenamefont {O'Donovan},
  \citenamefont {McBean}, \citenamefont {Jones}, \citenamefont {Ciston},
  \citenamefont {Scott},\ and\ \citenamefont {Ophus}}]{DaCosta2021}%
  \BibitemOpen
  \bibfield  {author} {\bibinfo {author} {\bibfnamefont {L.}~\bibnamefont
  {{Rangel DaCosta}}}, \bibinfo {author} {\bibfnamefont {H.~G.}\ \bibnamefont
  {Brown}}, \bibinfo {author} {\bibfnamefont {P.~M.}\ \bibnamefont {Pelz}},
  \bibinfo {author} {\bibfnamefont {A.}~\bibnamefont {Rakowski}}, \bibinfo
  {author} {\bibfnamefont {N.}~\bibnamefont {Barber}}, \bibinfo {author}
  {\bibfnamefont {P.}~\bibnamefont {O'Donovan}}, \bibinfo {author}
  {\bibfnamefont {P.}~\bibnamefont {McBean}}, \bibinfo {author} {\bibfnamefont
  {L.}~\bibnamefont {Jones}}, \bibinfo {author} {\bibfnamefont
  {J.}~\bibnamefont {Ciston}}, \bibinfo {author} {\bibfnamefont {M.~C.}\
  \bibnamefont {Scott}},\ and\ \bibinfo {author} {\bibfnamefont
  {C.}~\bibnamefont {Ophus}},\ }\bibfield  {title} {\bibinfo {title}
  {{Prismatic 2.0 – Simulation software for scanning and high resolution
  transmission electron microscopy (STEM and HRTEM)}},\ }\href
  {https://doi.org/10.1016/j.micron.2021.103141} {\bibfield  {journal}
  {\bibinfo  {journal} {Micron}\ }\textbf {\bibinfo {volume} {151}},\ \bibinfo
  {pages} {103141} (\bibinfo {year} {2021})}\BibitemShut {NoStop}%
\bibitem [{\citenamefont {Ophus}(2017)}]{Ophus2017}%
  \BibitemOpen
  \bibfield  {author} {\bibinfo {author} {\bibfnamefont {C.}~\bibnamefont
  {Ophus}},\ }\bibfield  {title} {\bibinfo {title} {{A fast image simulation
  algorithm for scanning transmission electron microscopy}},\ }\href
  {https://doi.org/10.1186/s40679-017-0046-1} {\bibfield  {journal} {\bibinfo
  {journal} {Adv. Struct. Chem. Imag.}\ }\textbf {\bibinfo {volume} {3}},\
  \bibinfo {pages} {1} (\bibinfo {year} {2017})},\ \Eprint
  {https://arxiv.org/abs/1702.01904} {arXiv:1702.01904} \BibitemShut {NoStop}%
\bibitem [{\citenamefont {Pryor}\ \emph {et~al.}(2017)\citenamefont {Pryor},
  \citenamefont {Ophus},\ and\ \citenamefont {Miao}}]{Pryor2017}%
  \BibitemOpen
  \bibfield  {author} {\bibinfo {author} {\bibfnamefont {A.}~\bibnamefont
  {Pryor}}, \bibinfo {author} {\bibfnamefont {C.}~\bibnamefont {Ophus}},\ and\
  \bibinfo {author} {\bibfnamefont {J.}~\bibnamefont {Miao}},\ }\bibfield
  {title} {\bibinfo {title} {{A streaming multi-GPU implementation of image
  simulation algorithms for scanning transmission electron microscopy}},\
  }\bibfield  {journal} {\bibinfo  {journal} {Adv. Struct. Chem. Imag.}\
  }\textbf {\bibinfo {volume} {3}},\ \href
  {https://doi.org/10.1186/s40679-017-0048-z} {10.1186/s40679-017-0048-z}
  (\bibinfo {year} {2017}),\ \Eprint {https://arxiv.org/abs/1706.08563}
  {arXiv:1706.08563} \BibitemShut {NoStop}%
\bibitem [{\citenamefont {Kingma}\ and\ \citenamefont {Ba}(2014)}]{Adamoptim}%
  \BibitemOpen
  \bibfield  {author} {\bibinfo {author} {\bibfnamefont {D.~P.}\ \bibnamefont
  {Kingma}}\ and\ \bibinfo {author} {\bibfnamefont {J.}~\bibnamefont {Ba}},\
  }\href {https://doi.org/10.48550/ARXIV.1412.6980} {\bibinfo {title} {Adam: A
  method for stochastic optimization}} (\bibinfo {year} {2014})\BibitemShut
  {NoStop}%
\bibitem [{\citenamefont {{Van Der Walt}}\ \emph {et~al.}(2014)\citenamefont
  {{Van Der Walt}}, \citenamefont {Sch{\"{o}}nberger}, \citenamefont
  {Nunez-Iglesias}, \citenamefont {Boulogne}, \citenamefont {Warner},
  \citenamefont {Yager}, \citenamefont {Gouillart},\ and\ \citenamefont
  {Yu}}]{Walt2014}%
  \BibitemOpen
  \bibfield  {author} {\bibinfo {author} {\bibfnamefont {S.}~\bibnamefont {{Van
  Der Walt}}}, \bibinfo {author} {\bibfnamefont {J.~L.}\ \bibnamefont
  {Sch{\"{o}}nberger}}, \bibinfo {author} {\bibfnamefont {J.}~\bibnamefont
  {Nunez-Iglesias}}, \bibinfo {author} {\bibfnamefont {F.}~\bibnamefont
  {Boulogne}}, \bibinfo {author} {\bibfnamefont {J.~D.}\ \bibnamefont
  {Warner}}, \bibinfo {author} {\bibfnamefont {N.}~\bibnamefont {Yager}},
  \bibinfo {author} {\bibfnamefont {E.}~\bibnamefont {Gouillart}},\ and\
  \bibinfo {author} {\bibfnamefont {T.}~\bibnamefont {Yu}},\ }\bibfield
  {title} {\bibinfo {title} {{Scikit-image: Image processing in python}},\
  }\href {https://doi.org/10.7717/peerj.453} {\bibfield  {journal} {\bibinfo
  {journal} {PeerJ}\ }\textbf {\bibinfo {volume} {2014}},\ \bibinfo {pages} {1}
  (\bibinfo {year} {2014})}\BibitemShut {NoStop}%
\bibitem [{\citenamefont {Pearson}(1901)}]{Pearson1901}%
  \BibitemOpen
  \bibfield  {author} {\bibinfo {author} {\bibfnamefont {K.}~\bibnamefont
  {Pearson}},\ }\bibfield  {title} {\bibinfo {title} {{LIII. On lines and
  planes of closest fit to systems of points in space}},\ }\href
  {https://doi.org/10.1080/14786440109462720} {\bibfield  {journal} {\bibinfo
  {journal} {Lond. Edinb. Dublin philos. mag. j. sci.}\ }\textbf {\bibinfo
  {volume} {2}},\ \bibinfo {pages} {559} (\bibinfo {year} {1901})}\BibitemShut
  {NoStop}%
\bibitem [{\citenamefont {Chen}\ and\ \citenamefont {Bovik}(2011)}]{Chen2011}%
  \BibitemOpen
  \bibfield  {author} {\bibinfo {author} {\bibfnamefont {M.~J.}\ \bibnamefont
  {Chen}}\ and\ \bibinfo {author} {\bibfnamefont {A.~C.}\ \bibnamefont
  {Bovik}},\ }\bibfield  {title} {\bibinfo {title} {{Fast structural similarity
  index algorithm}},\ }\href {https://doi.org/10.1007/s11554-010-0170-9}
  {\bibfield  {journal} {\bibinfo  {journal} {Journal of Real-Time Image
  Processing}\ }\textbf {\bibinfo {volume} {6}},\ \bibinfo {pages} {281}
  (\bibinfo {year} {2011})}\BibitemShut {NoStop}%
\end{thebibliography}%

\end{document}